\pdfoutput=1
\documentclass[prb,reprint,floatfix,superscriptaddress]{revtex4-2} 

\usepackage{amsmath,amsfonts,amssymb} %
\usepackage{mathtools} %
\ifx\phone\undefined
\usepackage{wasysym} 
\else
\let\Phone\phone
\let\phone\relax
\usepackage{wasysym}

\let\phone\Phone
\let\Phone\relax
\fi
\usepackage{bm} %
\usepackage[bbgreekl]{mathbbol} %
\usepackage{graphicx} %
\usepackage[dvipsnames,table]{xcolor} %
\usepackage{multirow,tabularx} %
\usepackage[acronym]{glossaries} %
\usepackage{braket} %
\usepackage[detect-weight=true]{siunitx} %
\usepackage{fancyhdr} %
\usepackage{microtype} %
\usepackage{natmove} 
\usepackage[caption=false]{subfig} %
\usepackage[breaklinks, %
colorlinks, linkcolor=Blue, citecolor=Blue, urlcolor=Blue]{hyperref} %
\usepackage[version=3]{mhchem} %
\usepackage{wrapfig} %
\AtBeginDocument{\usepackage{booktabs}}
\usepackage[linesnumbered,lined,boxed,ruled]{algorithm2e}

\DeclareSIUnit[number-unit-product = {\,}]\cal{cal}
\DeclareSIUnit\erg{erg}
\DeclareSIUnit\au{au}
\DeclareSIUnit\bohr{\text{\ensuremath{a}}_{0}}
\def\MyTitle{Optimization of the Qubit Coupled Cluster Ansatz on
  classical computers} %
\def\MyAuthora{Ilya G. Ryabinkin} %
\def\MyAuthorb{Seyyed Mehdi Hosseini Jenab} %
\def\MyAuthorc{Scott N. Genin} %
\def\MySubject{} %

\hypersetup{%
  pdftitle={\MyTitle{}}, %
  pdfauthor={\MyAuthora{}, \MyAuthorb{}, and \MyAuthorc{}}, %
  pdfsubject={\MySubject{}}, %
  pdfkeywords={} %
}

\graphicspath{{pics/}}

\newcolumntype{Y}{>{\centering\arraybackslash}X}

\newacronym[longplural={degrees of freedom},           %
            firstplural={degrees of freedom (DOF)},    %
            plural={DOF}]{DOF}{DOF}{degree of freedom} %

\newacronym[longplural={equations of motion},           %
            firstplural={equations of motion (EOM)},    %
            plural={EOM}]{EOM}{EOM}{equation of motion} %

\newacronym{TDSE}{TDSE}{time-dependent Schr\"odinger equation}
\newacronym{TIDSE}{TIDSE}{time-independent Schr\"odinger equation}
\newacronym{TDVP}{TDVP}{time-dependent variational principle} %
\newacronym{DFVP}{DFVP}{Dirac--Frenkel variational principle} %
\newacronym{MCTDH}{MCTDH}{multiconfguration time-dependent Hartree}
\newacronym{BCH}{BCH}{Baker--Campbell--Hausdorff}

\newacronym{vMCG}{vMCG}{variational multi-configurational Gaussian} %
\newacronym{FMS}{FMS}{full multiple spawning}                       %
\newacronym{G-MCTDH}{G-MCTDH}{Gaussian-based multi-configuration time-dependent Hartree} %
\newacronym{MAE}{MAE}{mean absolute error} %
\newacronym{pMAER}{\%MAER}{percentage of mean absolute error reduction} %

\newacronym[longplural={random-access memory},            %
            firstplural={random-access memory (RAM)},     %
            plural={RAM}]{RAM}{RAM}{random-access memory} %

\newacronym{GWP}{GWP}{Gaussian wavepacket}           %
\newacronym{TGWP}{TGWP}{thawed Gaussian wavepacket}  %
\newacronym{GHA}{GHA}{global harmonic approximation} %
\newacronym{LHA}{LHA}{local harmonic approximation}  %
\newacronym{ZPE}{ZPE}{zero-point energy}             %
\newacronym{VM}{VM}{vibrational mode}                %
\newacronym{CVM}{CVM}{curvilinear vibrational mode}  %

\newacronym{ML}{ML}{machine learning} %
\newacronym{ML-PES}{ML-PES}{machine-learned potential energy surface} %
\newacronym{KRR}{KRR}{kernel ridge regression}     %
\newacronym{GPR}{GPR}{Gaussian process regression} %

\newacronym{OLED}{OLED}{organic light-emitting diode}     %
\newacronym{NISQ}{NISQ}{noisy intermediate-scale quantum} %

\newacronym{JW}{JW}{Jordan--Wigner} %
\newacronym{BK}{BK}{Bravyi--Kitaev} %

\newacronym{QPE}{QPE}{quantum phase estimation}          %
\newacronym{VQE}{VQE}{variational quantum eigensolver}   %
\newacronym{QMF}{QMF}{qubit mean-field}                  %
\newacronym{QCC}{QCC}{qubit coupled cluster}             %
\newacronym{iQCC}{iQCC}{iterative qubit coupled cluster} %
\newacronym{iCCSDn}{iCCSDn}{iterative n-body excitation inclusive coupled-cluster single double} %
\newacronym{PQA}{PQA}{parametrized quantum annealing}    %
\newacronym{DIS}{DIS}{direct interaction set}            %
\newacronym[longplural={involutary linear combinations of anti-commuting Paulis}, %
            firstplural={involutory linear combinations of anti-commuting Paulis (ILCAP)}, %
            plural={ILCAP}]{ILCAP}{ILCAP}{involutory linear combination of anti-commuting Paulis} %
\newacronym{DHA}{DHA}{diagonal Hessian approximation} %

\newacronym{CAS}{CAS}{complete active space}    %
\newacronym{PES}{PES}{potential energy surface} %
\newacronym{PEC}{PEC}{potential energy curve}   %
\newacronym{AO}{AO}{atomic orbital}             %
\newacronym{MO}{MO}{molecular orbital}          %

\newacronym{CI}{CI}{configuration interaction}        %
\newacronym{FCI}{FCI}{full configuration interaction} %
\newacronym{CASCI}{CASCI}{complete active space configuration interaction} %
\newacronym{MCSCF}{MCSCF}{multiconfigurational self-consistent field} %
\newacronym{CASSCF}{CASSCF}{complete active space self-consistent field} %
\newacronym{CC}{CC}{coupled cluster}           %
\newacronym{UCC}{UCC}{unitary coupled cluster} %
\newacronym{GUCC}{GUCC}{generalized unitary coupled cluster} %
\newacronym{UCCSD}{UCCSD}{unitary coupled cluster singles and doubles} %
\newacronym{GUCCSD}{GUCCSD}{generalized unitary coupled cluster singles and doubles} %
\newacronym{CCSD}{CCSD}{coupled-cluster singles and doubles} %
\newacronym{CCSD-T}{CCSD(T)}{coupled-cluster singles and doubles and non-iterative triples} %
\newacronym{RHF}{RHF}{restricted Hartree--Fock}                     %
\newacronym{CIS}{CIS}{configuration interaction singles}            %
\newacronym{CISD}{CISD}{configuration interaction singles and doubles} %
\newacronym{ROHF}{ROHF}{restricted open-shell Hartree--Fock}        %
\newacronym{UHF}{UHF}{unrestricted Hartree--Fock}                   %
\newacronym{MPS}{MPS}{matrix product states}                        %
\newacronym{DMRG}{DMRG}{density-matrix renormalization group}       %
\newacronym{DFT}{DFT}{density-functional theory}                    %
\newacronym{TDDFT}{TDDFT}{time-dependent density-functional theory} %
\newacronym{ENPT}{ENPT}{Epstein-Nesbet perturbation theory}         %
\newacronym{MP}{MP}{M{\o}ller--Plesset perturbation theory}         %
\newacronym{MP2}{MP2}{second-order M{\o}ller--Plesset perturbation theory} %
\newacronym{MRMP2}{MRMP2}{second-order multi-reference M{\o}ller--Plesset perturbation theory} %

\newacronym{SQP}{SQP}{sequential quadratic programming} %
\newacronym{MMA}{MMA}{method of moving asymptotes}      %

\DeclareMathOperator{\vspan}{span} %
\DeclareMathOperator{\diag}{diag} %
\DeclarePairedDelimiter{\floor}{\lfloor}{\rfloor} %
\newcommand{\I}{\mathrm{i}\mkern1mu} %
\def\be{\begin{equation}} %
\def\ee{\end{equation}} %
\def\bea{\begin{eqnarray}} %
\def\eea{\end{eqnarray}} %

\begin{document}

\title{\MyTitle}

\author{\MyAuthora} %
\email{ilya.ryabinkin@otilumionics.com}

\author{\MyAuthorb}

\author{\MyAuthorc}
\affiliation{OTI Lumionics Inc., 3415 American Drive Unit~1, \\
  Mississauga, Ontario L4V\,1T4, Canada} %


\ifx\latin\undefined %
  \newcommand{\latin}{\textit} %
\fi%

\begin{abstract}
  Immense interest in quantum computing has prompted development of
  electronic structure methods that are suitable for quantum hardware.
  However, the slow pace at which quantum hardware progresses, forces
  researchers to implement their ideas on classical computers despite
  the obvious loss of any ``quantum advantage.'' As a result, the
  so-called \emph{quantum inspired} methods emerge. They allow one to
  look at the electronic structure problem from a different angle;
  yet, to fully exploit their capacity, efficient implementations are
  highly desirable. Here we report two schemes for improving the
  amplitude optimisation in the \gls{iQCC} method---a
  \acrlong{VQE}-type approach which is based on the \gls{QCC} Ansatz.
  Our first scheme approximates the \gls{QCC} unitary as a sum of
  symmetrical polynomials of generators up to a given order. The
  resulting energy expression allows for a flexible control of
  computational complexity via the order parameter. It also guaranties
  smoothness of trial energies and their derivatives, which is
  important for gradient-based optimization strategies. The second
  scheme limits the size of the expansion space in which the \gls{QCC}
  unitary is generated. It provides better control of memory
  requirements, but in general may lead to the non-smooth variation of
  energy estimates upon changes in amplitudes. It can be used,
  however, to extrapolate energies for a given set of amplitudes
  towards the exact \gls{QCC} value. Both schemes allow for a larger
  number of generators to be included into the \gls{QCC} form compared
  to the exact formulation. This reduces the number of iterations in
  the \gls{iQCC} method and/or leads to higher accuracy. We assess
  capabilities of the new schemes to perform \gls{QCC} amplitudes
  optimization for a few molecular systems: dinitrogen (\ce{N2}, 16
  qubits), water (\ce{H2O}, 36 qubits), and
  tris(2-(2,4-difluorophenyl)pyridine) iridium(III), (\ce{Ir(F2ppy)3},
  80 qubits).
\end{abstract}

\maketitle

\glsresetall

\section{Introduction}
\label{sec:introduction}

Quantum computing is a rapidly evolving area of computational science
that employs engineered quantum systems---quantum computers---to
perform calculations that are otherwise difficult for classical
computers~\cite{Feynman:1982/ijtp/467, Lloyd:1996/sci/1073,
  Ladd:2010/nature/45}. One of the promising applications of quantum
computing is finding the electronic
structure~\cite{Book/Helgaker:2000} of molecules and materials to
model their chemical and optical
properties~\cite{Marzari:2021/natmat/736}. Solving the electronic
structure problem amounts to finding eigenvalues---ground and excited
states---of the electronic Hamiltonian of a system of interest. The
first method proposed for solving this problem on quantum computers
was the \gls{QPE}
algorithm~\cite{Kitaev:1995/ArXiv/quant-ph/9511026,Cleve:1998/prsla/339,
  Abrams:1999/prl/5162, AspuruGuzik:2005/sci/1704,
  Lanyon:2010/nchem/106, Whitfield:2011/mp/735,
  OMalley:2016/prx/031007}. It quickly became clear, however, that
reliable estimate of eigenvalues with \gls{QPE} requires very complex
quantum circuits and long coherence times. For example, simulations of
the electronic structure of the iron-molybdenum cofactor (FeMoco)---an
active site of \ce{Mo}-dependent nitrogenase---were estimated to use
an astronomical ($10^{14}-10^{16}$) number of T-gate operations with
run time varying from a few months to several
years~\cite{Reiher:2017/pnas/7555-supp}. Subsequently, the
\gls{VQE}~\cite{Peruzzo:2014/ncomm/4213, Mcclean:2016/njp/023023,
  Kandala:2017/nature/242, Lee:2019/jctc/311, Romero:2018/qct/014008,
  Tilly:2022/pr/1} was suggested as a more frugal alternative to
\gls{QPE}. The \gls{VQE} method finds the ground electronic state of
an arbitrary qubit Hamiltonian
\begin{equation}
  \label{eq:ham}
  \hat H = \sum_i C_i \hat P_i,
\end{equation}
in which $C_i$ are numerical coefficients and
\begin{equation}
  \label{eq:P_def}
  \hat P_i = \hat\sigma_{i_1} \dots \hat\sigma_{i_q},\ 0 \le q \le (n-1) 
\end{equation}
are tensor products of Pauli elementary qubit operators
$\hat\sigma_j \in \{\hat x_j,\, \hat y_j,\, \hat z_j\}$ for $n$ qubits
(``Pauli words'' for short). In the context of the electric structure
problem, the Hamiltonian~\eqref{eq:ham} is derived from the
active-space electronic Hamiltonian of a molecular
system~\cite{Book/Helgaker:2000} using fermion-to-qubit mappings, such
as \gls{JW}, \gls{BK} or the more recent ternary-tree
mapping~\cite{Tranter:2018/jctc/5617, Zhang:2020/quantum/276}. Trial
energy
\begin{equation}
  \label{eq:qcc_energy}
  E(\mathbf{t}) = \braket{0|{\hat U^\dagger(\mathbf{t})}\hat H \hat
    U(\mathbf{t})|0}
\end{equation}
is evaluated on a quantum computer by measuring $\hat P_i$ in
Eq.~\eqref{eq:ham} on a state $\hat U(\mathbf{t}) \ket{0}$ of a
quantum register~\cite{Knill:2007/pra/012328, Chen:2013/pra/012109,
  Verteletskyi:2020/jcp/124114, Yen:2020/jctc/2400,
  Izmaylov:2020/jctc/190, Huggins:2021/npgqi/23} and summing up the
outcomes with coefficients $C_i$. Here, $\hat U(\mathbf{t})$ is a
parametrized unitary which must be realized as a quantum circuit and
$\ket{0}$ is a reference (initial) qubit state. Subsequently, the
measured trial energies $E(\mathbf{t})$ for different values of
parameters $\mathbf{t} = t_1, t_2, \dots$ are passed to a
\emph{classical} computer to find a minimum with respect to
$\mathbf{t}$. By the variational principle,
\begin{equation}
  \label{eq:emin}
  \min_{\mathbf{t}} E(\mathbf{t}) \ge E_0,
\end{equation}
where $E_0$ is the exact ground-state energy of the Hamiltonian
$\hat H$. We note that not only energies, but also energy gradients
can be measured using a parameter-shift
rule~\cite{Schuld:2019/pra/032331, Kottmann:2021/cs/3497,
  Izmaylov:2021/pra/062443}, which allows for using more efficient
gradient-based miminization techniques.

The parametrized unitary $\hat U(\mathbf{t})$ is a central quantity in
\gls{VQE}~\cite{Peruzzo:2014/ncomm/4213, Wecker:2015/pra/042303,
  Mcclean:2016/njp/023023, Kandala:2017/nature/242,
  Romero:2018/qct/014008, Ryabinkin:2018/jctc/6317,
  Grimsley:2018/nc/3007, Lee:2019/jctc/311,
  Lang:2020/arXiv/2002.05701, Yordanov:2021/cp/228,
  Freericks:2022/symmetry/494, Tilly:2022/pr/1}. Different flavours of
\gls{VQE} differ in the way how $\hat U(\mathbf{t})$ is constructed.
One specific form is the \gls{QCC}
Ansatz~\cite{Ryabinkin:2018/jctc/6317} inspired by the variational
approach introduced in Ref.~\citenum{Wecker:2015/pra/042303}. It is
defined as a product of exponents of generators $\hat T_k$,
\begin{equation}
  \label{eq:QCC_ans}
  \hat U(\mathbf{t}) = \prod_{k=1}^M \exp\left(-\I t_k\hat
    T_k/2\right),
\end{equation}
that are themselves Pauli words. Initially, it was proposed to screen
the entire space of all possible $\hat T_k$ whose dimensionality is
$4^n$, to include only generators that have non-zero values of the
first energy derivatives
\begin{equation}
  \label{eq:gradE}
  g_k = \frac{\mathrm{d} E[\hat T_k]}{\mathrm{d}t} =
  -\frac{\I}{2}\Braket{0|[\hat H, \hat T_k]|0},
\end{equation}
where $[\hat H, \hat T_k]$ is a commutator. It was later
recognized~\cite{Ryabinkin:2020/jctc/1055} that exponentially hard
screening can be replaced by a \emph{construction} of all such
generators given the qubit Hamiltonian, and this procedure has linear
complexity with respect to the number of terms in $\hat H$. If
$\hat H$ is a qubit image of the molecular electronic Hamiltonian, the
proposed algorithm produces qubit analogs of all single and double
excitations for a given set of molecular orbitals. The corresponding
Ansatz is, therefore, similar to the factorized
(``disentangled'')~\cite{Evangelista:2019/jcp/244112,
  Chen:2021/jctc/841, Xu:2023/symmetry/1429} form of a \gls{GUCC} wave
function~\cite{Lee:2019/jctc/311, Halder:2022/jcp/174117,
  Tribedi:2020/jctc/6317}.
Reference~\citenum{Ryabinkin:2020/jctc/1055} also introduces an
iterative scheme of building the \gls{QCC} form \emph{and}
transforming Hamiltonian, which allows for systematic convergence of
energy estimates to the exact ground-state energy of a given $\hat H$.
If all steps, including evaluation and optimization of trial
electronic energies~\eqref{eq:qcc_energy}, are done by a classical
computer, this \gls{iQCC} method becomes a quantum inspired algorithm.

The \gls{iQCC} method~\cite{Ryabinkin:2020/jctc/1055} can use as few
as one generator per iteration at expense of more iterations. Every
iteration produces a new dressed Hamiltonian
\begin{equation}
  \label{eq:dressed-H}
  \hat H^{(i)} = \left({\hat U^{(i)}(\mathbf{t}^{(i)}_\text{opt})}\right)^\dagger \hat H^{(i-1)}
  {\hat U^{(i)}(\mathbf{t}^{(i)}_\text{opt})},\  i=1, \ldots
\end{equation}
where $\hat H^{(i)}$ is the Hamiltonian at the $i$-th iteration,
$\hat H^{(0)}$ is the initial qubit-mapped active-space electronic
Hamiltonian, and ${\hat U^{(i)}(\mathbf{t}^{(i)}_\text{opt})}$ is a
\gls{QCC} unitary constructed at the $i$-th iteration evaluated at the
optimized values of amplitudes $\mathbf{t}^{(i)}_\text{opt}$. However,
dressing causes the Hamiltonian to quickly
expand~\cite{Lang:2021/jctc/66}, taxing CPU and memory of a classical
computer. Additionally, using only a small number of generators poses
a dilemma of artificial symmetry breaking, if not all generators that
preserve local or approximate symmetries can be included into the
Ansatz. Thus, enabling lengthier \gls{QCC} Ans\"atze improves
computational efficiency of the \gls{iQCC} method making it
competitive to the ``traditional'' quantum chemistry methods.

The current paper reports algorithm developments that allow as many as
hundred thousands of generators to be simultaneously optimized. We
propose two different approaches to simplify evaluation and
optimization of trial energies given by Eq.~\eqref{eq:qcc_energy}. In
the first one, we expand the \gls{QCC} unitary~\eqref{eq:QCC_ans} in a
series where terms are symmetric polynomials of generators and retain
all low-order terms up to a certain order. In the second, we
recursively construct a trial qubit state $\hat U(\mathbf{t})\ket{0}$
and drop the terms with smallest numerical coefficients to keep the
total number of them fixed.

We use three molecular systems in our study: dinitrogen (\ce{N2}, 16
qubits), water (\ce{H2O}, 36 qubits), and
\ce{Tris[2-(2,4-difluorophenyl)pyridine]Ir(III)} (\ce{Ir(F_2ppy)3}, 80
qubits)~\cite{Endo:2008/cpl/155, Genin:2022/ang/e202116175,
  Steiger:2024/arXiv/2404.10047}. For the first system we consider
molecular configurations with equilibrium \ce{N-N} distance (a weakly
correlated case) and a highly stretched geometry (a strongly
correlated case), as well as compute the entire \gls{PEC} to study the
convergence of energy and amplitudes in transition from one to
another. Water molecule is an illustration of an extrapolation
technique using the second scheme. Finally, the last system is used to
demonstrate scalability of our approach to large, industrially
relevant molecules.

\section{Theory}
\label{sec:theory}

\subsection{Symmetric-polynomial expansion of the \gls{QCC} unitary}
\label{sec:symm-polyn-expans}

Every exponent in the \gls{QCC} Ansatz~\eqref{eq:QCC_ans} can be
evaluated in a closed form as
\begin{equation}
  \label{eq:exp_iT}
  \exp(-\I t\hat T/2) = \cos\left(t/2\right) - \I\sin\left(t/2\right)\hat T,
\end{equation}
because any Pauli word is \emph{involutory},
$\hat T^2 = \mathcal{\hat E}$, where $\mathcal{\hat E}$ is the
identity operator. There are $2^M$ algebraic-independent terms in a
superposition which encodes the state $\hat U(\mathbf{t}) \ket{0}$ on
a quantum computer. This exponential complexity is a clear obstacle
for classical computers.

Expanding the product in the right-hand side of Eq.~\eqref{eq:QCC_ans}
using the identity~\eqref{eq:exp_iT} we obtain
\begin{widetext}
  \begin{align}
    \label{eq:qcc_expanded}
    \hat U(\mathbf{t}) = & \prod_{k=1}^M \cos\left(t_k/2\right) -
                           \I \sum_{j=1}^M \sin\left(t_j/2\right)\hat T_j\prod_{k\ne j}^M\cos\left(t_k/2\right) \nonumber \\
                         & - \sum_{i < j}^M\sin\left(t_i/2\right)\sin\left(t_j/2\right)\hat
                           T_i \hat T_j\prod_{k\ne
                           \{i,j\}}^M\cos\left(t_k/2\right)  + \dots
                           \nonumber \\
    = & \prod_{k=1}^M \cos\left(t_k/2\right) \left(1 -\I\sum_{j=1}^M
        \tan\left(t_j/2\right)\hat T_j  - \sum_{i < j}^M\tan\left(t_i/2\right)\tan\left(t_j/2\right)\hat
        T_i \hat T_j + \ldots \right)
  \end{align}
\end{widetext}
Equation~\eqref{eq:qcc_expanded} uses specific grouping of terms that
can be better explained on the following simple example. Consider a
unitary~\eqref{eq:QCC_ans} with $M = 3$ generators:
\begin{widetext}
  \begin{align}
    \label{eq:qcc_3t}
    \hat U(t_1, t_2, t_3) = {} &
                                 c_1c_2c_3 \quad \text{(0-th order)} \nonumber \\
                               & - \I \left(s_1c_2c_3\hat T_1 + c_1s_2c_3\hat T_2 + c_1c_2s_3\hat T_3 
                                 \right) \quad \text{(1-st order)} \nonumber \\
                               & - \left(s_1s_2c_3\hat T_1\hat T_2 +
                                 s_1c_2s_3\hat T_1\hat T_3 +
                                 c_1s_2s_3\hat T_2\hat T_3 \right) \quad
                                 \text{(2-nd order)} \nonumber \\
                               & + \I s_1s_2s_3\hat T_1\hat T_2\hat T_3
                                 \quad \text{(3-rd order)},
  \end{align}
\end{widetext}
where we use a short-hand notation $c_j = \cos\left(t_j/2\right)$,
$s_j = \sin\left(t_j/2\right)$. We collect and explicitly label terms
in Eq.~\eqref{eq:qcc_3t} by orders of the formal parameter $\lambda$
that can be attached to every $\hat T_j$ and then set to 1 at the end
of derivation~\footnote{Eq.~\eqref{eq:qcc_expanded} is, in fact, a
  \emph{generating function}.}. In Eq.~\eqref{eq:qcc_3t} one can
recognize that products of generators at each order are terms of the
elementary symmetric
polynomial~\cite[Ch.~1.2]{Book/Macdonald:1998symmetric} of that order.
For example, $\hat T_1\hat T_2$, $\hat T_1\hat T_3$, and
$\hat T_2\hat T_3$ are terms of the elementary symmetric polynomial of
the order 2 in three variables.

In a general case, Eq.~\eqref{eq:qcc_expanded}, there are $(M + 1)$
groups corresponding to orders $\lambda$ from 0 to $M$. Every
elementary symmetric polynomial of the order $K$ in $M$ variables
contains $\binom{M}{K} = \frac{M!}{K!(M-K)!}$ terms, and a sum of them
from $K=0$ to $M$ gives $2^M$ various products of $M$ generators,
again demonstrating the exponential complexity of the \gls{QCC} form.
Non-commutativity of generators has no consequences because of the
condition $i <j < k < \dots$ in the definition of the elementary
symmetric polynomials
(functions)~\cite[Ch.~1.2]{Book/Macdonald:1998symmetric}.

Grouping by orders of $\lambda$ can be further exploited. After the
factoring out the product of cosines of amplitudes
$\prod_{k=1}^M \cos\left(t_k/2\right)$ in Eq.~\eqref{eq:qcc_expanded},
the amplitude-dependent coefficients in parentheses are the terms of
elementary symmetrical polynomials in $\tan(t_k/2)$. If
$\forall k\, \tan(t_k/2) < 1$, at every order the products of
generators are multiplied by more quantities that are smaller than 1,
so that their numerical contributions diminish. Broadly defining a
weakly correlated regime as having all $\tan(t_k/2) < 1$, one may
expect that keeping only groups from $0$ to $K < M$ in
Eq.~\eqref{eq:qcc_expanded} provides a fair approximation to the full
expression.

The parenthesized expression in the last line of
Eq.~\eqref{eq:qcc_expanded} can be interpreted as a rank decomposition
of the operator-valued unitary
$\hat U(\mathbf{t}) = \hat U(t_1, t_2, \dots t_M)$ of the rank $M$ (as
it depends on $M$ ``indices'' $t_i$) as a sum of lower-rank operators
having the rank from $K=0$ to $M$. Rank decomposition is a powerful
tool for investigating complex quantum systems, which is behind such
approaches as the \gls{DMRG}~\cite{White:1992/prl/2863}, or, more
general, \gls{MPS}~\cite{Schollwock:2011/ap/92}, and Tree Tensor
Networks~\cite{Orus:2014/ap/117}.

Let us denote $\hat U^{[K]}(\mathbf{t})$ an approximation to the exact
$\hat U(\mathbf{t})$ in which all groups up to a $K$-th order in
$\lambda$ are kept. Then, an approximate energy functional \emph{could
  be} defined as
\begin{equation}
  \label{eq:W_def}
  W^{[K]}(\mathbf{t}) = \braket{0|(\hat
    U^{[K]})^\dagger(\mathbf{t}) \hat H \hat
    U^{[K]}(\mathbf{t})|0}. 
\end{equation}
However, in general, $W^{[K]}(\mathbf{t})$ is \emph{not} variationally
bound because $\hat U^{[K]}(\mathbf{t})$ is not unitary unless
$K = M$,
\begin{equation}
  \label{eq:u_norm}
  \left(U^{[K]}\right)^\dagger U^{[K]} \ne \mathcal{E},\ K < M.
\end{equation}
We anticipate poor quality of energies given by $W^{[K]}(\mathbf{t})$
functional unless enough terms are retained. The upper-boundary
property of Eq.~\eqref{eq:emin}, however, can be restored if we define
\begin{equation}
  \label{eq:E_K_def}
  E^{[K]}(\mathbf{t}) = \frac{\braket{0|(\hat
      U^{[K]})^\dagger(\mathbf{t}) \hat H \hat
      U^{[K]}(\mathbf{t})|0}}{\braket{0|(U^{[K]})^\dagger(\mathbf{t})
      U^{[K]}(\mathbf{t})|0}}, 
\end{equation}
which is our first, symmetric-polynomial approximate energy
functional. Its implementation is reported in
Appendix~\ref{sec:impl-ek_t}.

It must be emphasized that neither $W^{[K]}(\mathbf{t})$ nor
$E^{[K]}(\mathbf{t})$ coincide with an expression that follows from
the expansion of the \gls{QCC} \emph{energy},
Eq.~\eqref{eq:qcc_energy}, either in the commutator or in
perturbation-theory order~\cite{Cooper:2010/jcp/234102}.

\subsection{Limiting the expansion space}
\label{sec:limit-expans-space}

Formal algebraic independence of terms in Eq.~\eqref{eq:qcc_expanded}
does not imply linear independence of individual terms in the
\emph{vector} $\hat U(\mathbf{t})\ket{0}$. Every contribution to
$\hat U(\mathbf{t})\ket{0}$ has a form
$\kappa_{\bar q}(\mathbf{t})\hat T_{\bar q}\ket{0}$, where
$\hat T_{\bar q} = {\hat T_{q_1}} \cdots {\hat T_{q_r}},\ 1 \le r \le
M$ is a Pauli word and $\kappa_{\bar q}(\mathbf{t})$ are functions of
amplitudes. Any $\hat T_{\bar q}$ can be uniquely factorized as
\begin{equation}
  \label{eq:pauli_factorization}
  {\hat T}_{\bar q} = \phi_{\bar q} {\hat X}_{\bar q} {\hat Z}_{\bar q},
\end{equation}
where $\phi_{\bar q} = \{\pm 1,\, \pm\I\}$ is a factorization phase,
and ${\hat X}_{\bar q}$ and ${\hat Z}_{\bar q}$ are some Pauli words
that are products of only $\hat x$ or $\hat z$ Pauli elementary
operators [\latin{cf.} Eq.~\eqref{eq:P_def}], respectively. Because
our reference state $\ket{0}$ is an eigenvector of
${ \hat Z}_{\bar q}$ for any $\bar q$:
${\hat Z}_{\bar q}\ket{0} = f_{\bar q}\ket{0}$, $f_{\bar q} = \pm 1$,
we can write
\begin{equation}
  \label{eq:T_ket0}
  \hat T_{\bar q}\ket{0} = \phi_{\bar q} \hat X_{\bar q} \hat Z_{\bar
    q} \ket{0} = \phi_{\bar q} f_{\bar q} \hat X_{\bar q} \ket{0}.
\end{equation}
Vectors (``excited states'') $\hat X \ket{0}$ are linearly independent
for different $\hat X$. It may occur, however, that different products
of generators (\latin{i.e.} with different $\bar q$) factorize to the
\emph{same} $\hat X$ operator, thus producing identical vectors albeit
with different phases and different $\kappa(\mathbf{t})$. This
coalescence of terms inevitably happen if $M > n$ as it is not
possible to generate more than $2^n$ linearly independent vectors in
the Hilbert space of dimensionality $2^n$ (for $n$ qubits).


The idea of approximating the vector $\hat U(\mathbf{t})\ket{0}$
rather than the \gls{QCC} unitary comes out naturally. This vector
lives in a $2^n$-dimensional Hilbert space and evaluating and storing
it on a classical computer is impossible for sufficiently large $n$
and $M$. Hence, confining it in an $N$-dimensional subspace,
$N \ll 2^n$, would be highly desirable. The main challenge is how to
find a basis of that subspace, which, for example, maximizes the
projection of $\hat U(\mathbf{t})\ket{0}$ onto it.

Consider a step-by-step application of the unitary~\eqref{eq:QCC_ans}
to the reference vector $\ket{0}$~\footnote{We use lowercase bold
  letters to denote vectors in $2^n$ qubit Hilbert space and capital
  bold letters to define subspaces where those vectors live in.}.
Initially we have $\mathbf{v}_0 = \ket{0}$ and
$\mathbf{V}_0 = \vspan\{\ket{0}\}$. Obviously,
$\mathbf{v}_0 \in \mathbf{V}_0$. Then, applying exponents from the
\gls{QCC} unitary~\eqref{eq:QCC_ans} starting from the rightmost
factor $\exp{\left(-\I t_M \hat T_M/2\right)}$, we have:
\begin{widetext}
  \begin{align}
    \label{eq:v1}
    \mathbf{v}_1 & = \cos(t_M/2)\mathbf{v}_0 -\I\sin(t_M/2)\, \hat T_M\mathbf{v}_0, \\ \nonumber 
    \mathbf{V}_1 & = \vspan\{\ket{0},  -\I {\hat T}_M\ket{0}\} \\
    \mathbf{v}_2 & = \cos(t_{M-1}/2)\mathbf{v}_1  -\I\sin(t_{M-1}/2) \hat T_{M-1}\mathbf{v}_1, \\ \nonumber
    \mathbf{V}_2 & = \vspan\{\ket{0},  -\I {\hat T}_M\ket{0}, -\I {\hat T}_{M-1}\ket{0}, -{\hat T_{M-1}} {\hat T}_M\ket{0}\} \\  \nonumber
                 & \ldots  \\
    \mathbf{v}_k & = \cos(t_{M-k+1}/2)\mathbf{v}_{k-1}  -\I\sin(t_{M-k+1}/2) \hat T_{M-1}\mathbf{v}_{k-1}, \\ \nonumber
    \mathbf{V}_k & = \vspan\{\ket{0},  -\I {\hat T}_M\ket{0}, \ldots -\I
                   {\hat T}_{M-k+1}\ket{0}, \ldots, (-\I)^k\prod_{j=k-1}^{0}
                   {\hat T}_{M-j}\ket{0}\} \\  \nonumber
                 & \quad 2 < k \le M \nonumber
  \end{align}
\end{widetext}
Here we use \gls{QCC} generators multiplied by the factor $(-\I)$
because those products have purely real matrix elements in the
computational basis set.

If no truncation is done up to the $k$-th level and no redundancy
occurs, then $\dim{\mathbf{V}_k} = 2^k$. If, additionally, $k = M$
then $\dim{\mathbf{V}_M} = 2^M$ recovering the exponential complexity
of the \gls{QCC} unitary with respect to the number of generators.
During this \emph{expansion} stage the action of every exponential
factor in the \gls{QCC} unitary is represented exactly.

To suppress exponential growth the following strategy is followed:
when the dimensionality of $\mathbf{V}_k$ exceeds a pre-defined number
$N$, then $(\dim{\mathbf{V}_k} - N)$ basis vectors which have the
smallest absolute coefficients in $\mathbf{v}_k$ are discarded,
limiting the dimensionality of $\mathbf{V}_k$ by $N$. After the
application of the next exponential factor the size of
$\mathbf{V}_{k+1}$ cannot grow for more than $2N$ (less if a
coalescence of terms occurs), and analysis of $\mathbf{v}_{k+1}$ is
repeated limiting the dimensionality of $\mathbf{V}_{k+1}$ by $N$ as
well. This is the \emph{compression} stage. The process continues
until all exponents in the \gls{QCC} unitary are exhausted. At the
end, one arrives at the final composition of the $\mathbf{V}_M$ with
$\dim \mathbf{V}_M \le N$ and finds the expansion coefficients of
$\mathbf{v}_M$ that are, in general, functions of all amplitudes.

If truncation happened at the $k$-th level, then the truncated vector
$\mathbf{v}_k$ has the norm less than one, $\|\mathbf{v}_k\| < 1$.
This breaks the unitarity of the \gls{QCC} Ansatz~\eqref{eq:QCC_ans}.
We recover it by re-normalizing $\mathbf{v}_k$ after truncating by
re-scaling its coefficients by $1/\|\mathbf{v}_k\|$. These
re-normalization factors can be collected and stored to estimate the
norm loss of the $\hat U(\mathbf{t})\ket{0}$ during truncation.

Once properly normalized $\mathbf{v}_M$ is computed, the approximate
energy corresponding to the set of amplitudes
$\mathbf{t} = (t_1, t_2, \dots, t_n)$ can be evaluated as
\begin{equation}
  \label{eq:E_N_def}
  F^{[N]}(\mathbf{t}) = \braket{\mathbf{v}_M|H|\mathbf{v}_M}
\end{equation}
where expansion coefficients (``coordinates'') of $\mathbf{v}_M$ in
$\mathbf{V}_M$ carry the information about the amplitudes.
Equation~\eqref{eq:E_N_def} introduces our second approximate energy
functional, $F^{[N]}(\mathbf{t})$. Its implementation is reported in
Appendix~\ref{sec:impl-fn_t}.

\subsection{General properties of the proposed energy functionals}
\label{sec:gener-prop-prop}

Accuracy of $E^{[K]}(\mathbf{t})$, Eq.~\eqref{eq:E_K_def}, and
$F^{[N]}(\mathbf{t})$, Eq.~\eqref{eq:E_N_def}, is controlled by the
order parameter $K \le M$ and the expansion space dimensionality
$N \le 2^{\min\{M,n\}}$, respectively. CPU and storage requirements
for evaluating $\hat U^{[K]}(\mathbf{t})$ rises steeply with $K$ and
$M$ as
\begin{equation}
  \label{eq:N_K_for_E_K}
  N(M, K) = \sum_{k=0}^K \binom{M}{k} = \sum_{k=0}^{\floor{K/2}} \binom{M+1}{K-2k}.
\end{equation}
The last equality gives compact explicit expressions for $K=1$ and
$2$: $(M+1)$ and $\frac{M(M+1)}{2} + 1$, respectively. Numerical
trends for other $M$ and $K$ are illustrated by
Table~\ref{tab:U_K_size}.
\begin{table}
  \centering
  \caption{The number of terms in $\hat U^{[K]}(\mathbf{t})$ as a
    function of $K$ and $M = \dim{\mathbf{t}}$. In the limiting case
    $M = K$, the number of terms is $2^M$; if $K \ll M$ it is
    $O(M^K)$.}
  \begin{tabularx}{1.0\linewidth}{l|XXXXX}
    \toprule
    M\textbackslash K \quad& 1  & 2   & 5     & 10     &  20 \\
    \midrule
    1   & \num{2}  & -         & -           & -            &  -      \\
    2   & \num{3}  & \num{4}   & -           & -            &  -      \\
    5   & \num{6}  & \num{16}  & \num{32}    & -            &  -      \\
    10  & \num{11} & \num{56}  & \num{638}   & \num{1024}   &  -      \\
    20  & \num{21} & \num{211} & \num{21700} & \num{616666} & \num{1048576} \\
    \bottomrule
  \end{tabularx}
  \label{tab:U_K_size}
\end{table}
$E^{[K]}(\mathbf{t})$ provides energies and derivatives that are
smooth with respect to variation of amplitudes. Contrary to that,
$F^{[N]}(\mathbf{t})$ is, in general, \emph{non-differentiable} as the
composition of the truncated space can be drastically different for
close but distinct values of input amplitudes, leading to non-smooth
variation of energies with respect to $\mathbf{t}$, with an exception
of the limiting case $N = 2^{\min\{M,n\}}$. Thus,
$F^{[N]}(\mathbf{t})$ is less suitable for amplitude optimization than
$E^{[K]}(\mathbf{t})$.

\subsection{\Gls{DHA} and new generator ranking}
\label{sec:diag-hess-appr}

$E^{[K]}(\mathbf{t})$ for $K=1$ can be simplified further. Consider
\begin{align}
  \label{eq:U_1}
  \hat U^{[1]}(\mathbf{t}) = & \prod_{k=1}^M \cos\left(t_k/2\right)
                               \left(1 - \I\sum_{j=1}^M
                               \tan\left(t_j/2\right)\hat T_j\right)
                               \nonumber \\
  {} = & \mathcal{R}_M(\mathbf{t}) \left(1 - \I\sum_{j=1}^M
         C_j\hat T_j\right),
\end{align}
where we defined
\begin{align}
  \label{eq:R_def}
  \mathcal{R}_M(\mathbf{t}) = & \prod_{k=1}^M \cos\left(t_k/2\right),
  \\
  \label{eq:C_def}
  C_j = & \tan\left(t_j/2\right).
\end{align}
From Eq.~(\ref{eq:E_K_def}) $E^{[1]}(\mathbf{t})$ is then
\begin{widetext}
  \begin{align}
    \label{eq:E_1}
    E^{[1]}(\mathbf{t}) = & \left(1 +
                            \sum_{k=1}^MC_k^2\right)^{-1}\left(1 -
                            \I\sum_{k=1}^MC_k\braket{0|[\hat H, \hat
                            T_k]|0} + \sum_{j,\,k=1}^M
                            C_jC_k\braket{0|\hat T_j \hat H \hat
                            T_k|0}\right) \nonumber \\
    {} = & \left(1 + \sum_{k=1}^MC_k^2\right)^{-1}\left(1 +
           2\sum_{k=1}^M C_kg_k +
           \sum_{j,\,k=1}^M C_jC_k\braket{0|\hat T_j \hat H \hat
           T_k|0}\right),
  \end{align}
\end{widetext}
because $\braket{0|\hat T_j \hat T_k|0} = \delta_{jk}$ (the Kronecker
delta). We also used the definition of \gls{QCC} energy gradients
$g_k$, Eq.~\eqref{eq:gradE}.

Minimization of $E^{[1]}(\mathbf{t})$ with respect to
$\mathbf{C} = C_1, C_2, \ldots$ [see Eq.~\eqref{eq:C_def}] leads to
the standard eigenvalue problem for the \emph{Hessian} matrix
$\mathbf{H}$,
\begin{equation}
  \label{eq:Hess_eigenproblem}
  \mathbf{H}\mathbf{C} = E\mathbf{C}.
\end{equation}
Minimization with respect to amplitudes $\mathbf{t}$ rather than
$\mathbf{C}$ by demanding
\begin{equation}
  \label{eq:U_1_min}
  \frac{\partial E^{[1]}(\mathbf{t})}{\partial t_j} = \sum_k
  \frac{\partial E^{[1]}(\mathbf{C})}{\partial C_k} \frac{\partial
    C_k}{\partial t_j} = 0 
\end{equation}
leads to the \emph{same} minimum, since the Jacobian of
transformation,
$\left\{\frac{\partial C_k}{\partial t_j}\right\} =
\diag{\left\{\frac{\partial C_j}{\partial t_j}\right\}} =
\diag{\left\{\frac{1}{2\cos^2\left(t_j/2\right)}\right\}}$ is singular
only if $\cos\left(t_j/2\right) = 0$ for some $j$. In this case, one
of the rotated states, $\hat T_j\ket{0}$, [see Eq.~\eqref{eq:exp_iT}]
is orthogonal to the reference state $\ket{0}$, which implies
reference \emph{instability}. We assume this situation never occurs.

The \acrfull{DHA} is introduced by ignoring all the off-diagonal
($j \ne k$) terms in the last sum in Eq.~(\ref{eq:E_1}), which implies
the following structure of the $(M+1)\times(M+1)$ Hessian matrix:
\begin{equation}
  \label{eq:HA}
  \mathbf{H} =
  \begin{pmatrix}
    E_0    & g_1    & g_2    & \cdots  &  g_M   \\
    g_1    & E_1    & 0      & \cdots  &  0     \\
    g_2    & 0      & E_2    &         &  0     \\ 
    \vdots & \vdots &        & \ddots  & \vdots \\
    g_M    & 0      & \multicolumn{2}{c}{\ldots}  & E_M  
  \end{pmatrix},
\end{equation}
where $E_0 = \braket{0|\hat H|0}$ is the reference energy, $g_k$ are
\gls{QCC} gradients, and
\begin{equation}
  \label{eq:QCC_ex_refs}
  E_k = \braket{0|\hat T_k\hat H \hat T_k|0} =
  \braket{0|\hat X_k\hat H \hat X_k|0}, \ k = 1 \dots M
\end{equation}
are the excited-state energies. Equation~(\ref{eq:QCC_ex_refs}) uses
Eq.~\eqref{eq:T_ket0} to get rid of phase factors and refer to unique
basis states $\hat X_k\ket{0}$.

The ground-state eigenvalue (energy) $E$ and the corresponding
amplitudes $\mathbf{t} = 2\arctan{(\mathbf{C})}$ for the
eigenproblem~\eqref{eq:Hess_eigenproblem} with the
matrix~(\ref{eq:HA}) can be found via an iterative procedure. In fact,
$E$ is identical to the energy given by the Brillouin--Wigner
perturbation theory~\cite{Brillouin:1932/jpr/373,
  Wigner:1937/collection/131} for the state $\ket{0}$.

Because the optimal amplitudes in \gls{DHA} can be efficiently found
even for $\dim{\mathbf{t}} = M \gtrapprox 10^{8}$, the absolute values
$|t_k|$ can be used for \emph{ranking} generators---to include the
top-ranked ones into the \gls{QCC} Ansatz. This is a new alternative
to the ranking schemes proposed in
Ref.~\citenum{Ryabinkin:2021/qst/024012}, see
Sec.~\ref{sec:generator-ranking}.

\section{Simulations details}
\label{sec:simulations}

\subsection{General setup}
\label{sec:gener-char-trial}

The active-space fermionic Hamiltonians for these molecules are
written in the spin-orbital basis and converted to a qubit form using
the \gls{JW} transformation. Spin-orbitals are grouped as
$\phi_1\ket{\alpha}$, $\phi_1\ket{\beta}$, $\phi_2\ket{\alpha}$,
\latin{etc}, where $\phi_i$ are the \gls{RHF} \emph{spatial} orbitals
and $\ket{\alpha}$ and $\ket{\beta}$ are spin-1/2 eigenfunctions.
\Gls{RHF} \glspl{MO} generation and the \gls{AO} to \gls{MO}
transformation for the active spaces were done using the modified
version of the \textsc{gamess}~\cite{gamessus, gamessus-2} program
suit. Active-space \gls{CISD}, \gls{CCSD}, and \gls{CASCI}
calculations for the \ce{N2} molecule were done in
Psi4~\cite{Psi4-1.1} program package, version 1.3.1, while \gls{CISD},
CISDT, CISDTQ, \gls{CCSD} and \gls{CCSD-T} active-space energies for
\ce{Ir(F2ppy)3} were calculated in \textsc{gamess}. Preparing qubit
Hamiltonians and iQCC iterations are performed by a proprietary suit
of programs written in \textsc{julia}~\cite{Julia:2017} language.
Energy minimization was always performed with \textsc{l-bfgs}
algorithm~\cite{lbfgs-scipy-Byrd:1995/jsc/1190,
  lbfgs-scipy-Zhu:1997/atms/550} implemented in the
\texttt{NLopt}~\cite{NLOpt} \textsc{julia} package.

\subsection{Generator ranking}
\label{sec:generator-ranking}

Generators included into the \gls{QCC} Ansatz were ranked by the
formula~\cite{Ryabinkin:2021/qst/024012}:
\begin{equation}
  \label{eq:un1_ranking}
  r_j = \left|\arcsin\frac{2g_i}{\sqrt{D_i^2 + 4g_i^2}}\right| =
  \left|\arctan{\frac{2g_i}{D_i}}\right|,
\end{equation}
where $g_i$ is the gradient associated with a candidate generator
$\hat T_i$ [see Eq.~\eqref{eq:gradE}] and
$D_i = \braket{0|\hat H|0} - \braket{0|\hat T_i \hat H \hat T_i|0} =
E_0 - E_i$ is a difference between the reference energy $E_0$ and the
excited state energies $E_i$, see Eq.~(\ref{eq:QCC_ex_refs}). In the
limit of small couplings $g_i \ll D_i$ this formula is reduced to the
Epstein-Nesbet~\cite{Epstein:1926/pr/695, Nesbet:1955/prsl/312}
perturbation-theory expression for the first-order absolute
amplitudes, $|{2g_i}/{D_i}|$. Equation~(\ref{eq:un1_ranking}),
however, provides robust ranking even in the case of strong
correlation when $D_i \ll g_i$, and a perturbation theory fails.

\section{Results and discussion}
\label{sec:results-discussion}

\subsection{Dinitrogen (\ce{N2}, 16 qubit)}
\label{sec:n2}

For the \ce{N2} molecule we performed \gls{RHF} calculations using
Dunning's cc-pVDZ basis set~\cite{Dunning:1989/jcp/1007}. The
\gls{CAS} included all 8 valence (comprised of 2s and 2p orbitals of
both \ce{N} atoms) \gls{RHF} \glspl{MO} and 10 valence electrons; it
is CAS(10,~8). The 16-qubit Hamiltonian was produced as described in
Sec.~\ref{sec:simulations}. It contains 825 terms larger than
$\num{e-8}$ in magnitude and has 136 generators, which correspond to
single and double fermionic excitations, with potentially non-zero
gradients~(\ref{eq:gradE}). Some of them, however, have numerically
zero gradients for the closed-shell Hartree--Fock reference with 8
electrons, for example those that correspond to the occupied-occupied
or virtual-virtual excitations. On the other hand, including them to
the \gls{QCC} Ansatz makes it similar to the disentangled form of
\gls{GUCCSD} method~\cite{Lee:2019/jctc/311} or a unitarized version
of the \gls{iCCSDn} theory~\cite{Halder:2022/jcp/174117}.

\begin{table*}
  \centering
  \caption{\ce{N2} molecule with \ce{d(N-N)} = \SI{2.118}{\bohr}. The
    optimal energies $E^{[K]}_\text{opt}$, exact energies at optimized
    amplitudes $E(\mathbf{t}^{[K]}_\text{opt})$, and the norm of
    deviation
    $\left\Vert \mathbf{t}^{[K]}_\text{opt} - \mathbf{t}_\text{exact}
    \right\Vert$ along with the number of terms in
    $\hat U^{[K]}(\mathbf{t})$ and the length (dimensionality) of
    $\hat U^{[K]}(\mathbf{t})\ket{0}$ for $K$ from 0 to $M = 22$.
    $\mathbf{t}_\text{exact}$ are the amplitudes optimized for
    $K = M$. A special case $K=0$ corresponds to the amplitudes
    optimized using \gls{DHA}, see Sec.~\ref{sec:diag-hess-appr}.}
  \begin{tabularx}{1.0\textwidth}{@{}lllXXl@{}}
    \toprule
    $K$& Terms in $\hat U^{[K]}(\mathbf{t})$ & $\dim{(\hat U^{[K]}(\mathbf{t})\ket{0})}$ & $E^{[K]}_\text{opt}$ & $E(\mathbf{t}_\text{opt})$ & $\left\Vert \mathbf{t}_\text{opt} - \mathbf{t}_\text{exact} \right\Vert$ \\
    \midrule
    0  &       23  &   23 & $-109.058174408$ & $-109.011843381$ & \num{2.25E-01} \\
    1  &       23  &   23 & $-109.027206664$ & $-109.027501062$ & \num{6.59E-02} \\
    2  &      254  &  146 & $-109.028505327$ & $-109.028475415$ & \num{5.66E-03} \\
    3  &     1794  &  438 & $-109.028482438$ & $-109.028483240$ & \num{6.06E-04} \\
    4  &     9109  &  764 & $-109.028483783$ & $-109.028483322$ & \num{4.76E-05} \\
    5  &    35443  &  956 & $-109.028483312$ & $-109.028483322$ & \num{1.69E-06} \\
    6  &   110056  & 1016 & $-109.028483323$ & $-109.028483322$ & \num{1.01E-07} \\
    7  &   280600  & 1024 & $-109.028483322$ & $-109.028483322$ & \num{7.06E-08} \\
    8--22 & Eq.~(\ref{eq:N_K_for_E_K}) & 1024 & $-109.028483322$ & $-109.028483322$ & \num{7.19E-08} \\
    Exact  &       &      &                & $-109.028483322$ &  \num{0}        \\
    \bottomrule
  \end{tabularx}
  \label{tab:N2_at2.118_nent_22}
\end{table*}
First, we study the convergence of $E^{[K]}$, Eq.~(\ref{eq:E_K_def}),
with respect to $K$ for the case of weak correlation. The weakly
correlated case is represented by the \ce{N2} molecule with
$\ce{d(N-N)} = \SI{2.118}{\bohr}$, which is close to the equilibrium
geometry at the \gls{FCI} level in the cc-pVDZ basis. At this
internuclear distance the singlet ground state of the molecule is
well-separated from higher-spin states, and \gls{iQCC} iterations
converge eventually to the exact ground-state singlet energy. We
consider only the first iteration when $E^{[K]}$ is optimized using
the fermionic Hamiltonian $\hat H^{(0)}$ [\latin{c.f.}
Eq.~\eqref{eq:dressed-H}]. Generators were ranked by
Eq.~(\ref{eq:un1_ranking}), and we selected $M = 22$ of them; this
value is close to the limit of optimization of un-truncated
$\hat U(\mathbf{t})$ due to CPU and memory limitations of a typical
workstation.

The optimized energies $E^{[K]}_\text{opt}$, exact energies at the
optimized amplitudes $E(\mathbf{t}^{[K]}_\text{opt})$, the norm of
deviation
$\left\Vert \mathbf{t}^{[K]}_\text{opt} - \mathbf{t}_\text{exact}
\right\Vert$ along with the number of terms in
$\hat U^{[K]}(\mathbf{t})$ and the length (dimensionality) of
$\hat U^{[K]}(\mathbf{t})\ket{0}$ for $K=1$ to $M$ are shown in
Table~\ref{tab:N2_at2.118_nent_22}. $\mathbf{t}_\text{exact}$ are the
amplitudes optimized for $K = M$. A special case labelled by $K=0$
corresponds to the optimization in the \gls{DHA} approximation, see
Sec.~\ref{sec:diag-hess-appr}. As follows from the
Table~\ref{tab:N2_at2.118_nent_22}, the optimized energies and
amplitudes converge quickly with increasing of $K$: sub-millihartree
accuracy is already achieved at $K = 2$, and a numerically converged
result can be obtained for $K=6$. The norm of deviation
$\left\Vert \mathbf{t}^{[K]}_\text{opt} - \mathbf{t}_\text{exact}
\right\Vert$ decreases roughly by an order of magnitude each time $K$
increases by 1 until $K=7$, and then stalls because the numerical
precision of the objective function (energy) is limited by its
float-point representation on a computer. Overall, these results
emphasizes the fact that in a weakly correlated system the exact
representation of a system wave function is highly redundant because
of the dominance of vanishing numerical
contributions~\cite{Ivanic:2001/tca/339}.

Secondly, we investigate convergence of of $E^{[K]}(\mathbf{t})$,
Eq.~(\ref{eq:E_K_def}), with respect to $K$ for the case of strong
correlation represented by a stretched \ce{N2} molecule with
\ce{d(N-N)} = \SI{4.0}{\bohr}. At this nuclear configuration the
singlet ground state is highly multiconfigurational. The \gls{CI}
coefficient of the Hartree--Fock reference is approximately \num{0.5}.
\Gls{iQCC} iterations in this case converge to a superposition of
ground and higher-multiplicity states. Fundamentally, this happens
because some of generators are not spin-adapted and contribute to the
creation of both open-shell singlet and $S_z = 0$ components of
triplets. In this case energy optimization suffers from strong spin
contamination and, unless constrained optimization techniques or
penalties~\cite{Ryabinkin:2019/jctc/249, Genin:2022/ang/e202116175}
are employed, the spin symmetry breaking is almost unavoidable.
Ultimately, this is reminiscent L\"owdin symmetry
dilemma~\cite{Lowdin:1963/rmp/496}, which plagued the Hartree--Fock
optimization of singlet states in the spin-unrestricted formalism. For
the sake of demonstration, we did \emph{not} add penalty operators to
the Hamiltonian, but tracked the mean value of $\hat S^2$ operator
computed at the set of optimized amplitudes. Additionally, in this
case we used $M = 19$ to avoid cutting the group of generators with
almost identical values of ranking---such a precaution measure is
necessary to prevent \emph{artifical} symmetry breaking.

The convergence of $E^{[K]}_\text{opt}$ and related quantities are
reported in Table~\ref{tab:N2_at4.0_nent_19}.
\begin{table*}
  \centering
  \caption{\ce{N2} molecule with $\ce{d(N-N)} = \SI{4.0}{\bohr}$. The
    optimal energies $E^{[K]}_\text{opt}$, exact energies at optimized
    amplitudes $E(\mathbf{t}^{[K]}_\text{opt})$, the norm of deviation
    $\left\Vert \mathbf{t}^{[K]}_\text{opt} -
      \mathbf{t}_\text{exact}\right\Vert$, and the expectation value
    for the total spin-squared operator
    $\Braket{\hat S^2(\mathbf{t}_\text{opt})}$ along with the number
    of terms in $\hat U^{[K]}(\mathbf{t})$ and the length
    (dimensionality) of $\hat U^{[K]}(\mathbf{t})\ket{0}$ for $K$ from
    0 to $M = 19$. $\mathbf{t}_\text{exact}$ are the amplitudes
    optimized for $K = M$. A special case $K=0$ corresponds to the
    amplitudes optimized using \gls{DHA}, see
    Sec.~\ref{sec:diag-hess-appr}.}
  \label{tab:N2_at4.0_nent_19}
  \begin{tabularx}{1.0\textwidth}{@{}lllXXcc@{}}
    \toprule
    $K$& Terms in $\hat U^{[K]}(\mathbf{t})$ & $\dim{\hat U^{[K]}(\mathbf{t})\ket{0}}$ & $E^{[K]}_\text{opt}$ & $E(\mathbf{t}_\text{opt})$ & $\left\Vert \mathbf{t}_\text{opt} - \mathbf{t}_\text{exact} \right\Vert$ & $\Braket{\hat S^2(\mathbf{t}_\text{opt})}$ \\
    \midrule
    0   & 20	&  20 & -108.8586872116&  -108.5895989786&  \num{2.4E+00}&  0.065012 \\
    1   & 20    &  20 & -108.6050091792&  -108.1867637296&  \num{2.3E+00}&  0.820330 \\
    2   & 191   &  87 & -108.6961743172&  -108.6242746563&  \num{8.3E-01}&  2.162786 \\
    3   & 1160  &  128& -108.7219620714&  -108.7152274369&  \num{4.6E-01}&  2.068216 \\
    4   & 5036  &  128& -108.7266865097&  -108.7261532367&  \num{2.6E-01}&  2.082183 \\
    5   & 16664 &  128& -108.7268735613&  -108.7267206541&  \num{3.9E-02}&  2.144843 \\
    6   & 43796 &  128& -108.7267755035&  -108.7267551552&  \num{5.1E-03}&  2.154885 \\
    7   & 94184 &  128& -108.7267564062&  -108.7267562398&  \num{8.0E-04}&  2.153763 \\
    8   & 169766&  128& -108.7267564209&  -108.7267562538&  \num{6.9E-05}&  2.153521 \\
    9   & 262144&  128& -108.7267562453&  -108.7267562539&  \num{3.5E-06}&  2.153535 \\
    10--19 & Eq.~(\ref{eq:N_K_for_E_K}) &  128& -108.7267562539&  -108.7267562539&  $O(10^{-7})$&  2.153534 \\
   Exact&       &     &                &  -108.7267562539&    \num{0}       & 2.153534 \\ 
   \bottomrule
  \end{tabularx}
\end{table*}
Unsurprisingly, $E^{[K]}$ struggles to obtain the correct solution for
$K$ up to 3. Starting from $K = 4$ onward, amplitudes converge
geometrically with increasing $K$, lowering the norm of the difference
with the exact amplitudes by roughly one order of magnitude per unit
step. It is clear also that optimization without constraints/penalties
has a tendency to converge to the high-spin solution rather than a
singlet---this can be seen from expectation values of the total
spin-squared operator $\hat S^2$. Energy is converged to
sub-millihartree accuracy at $K=4$, but the mean value of $\hat S^2$
roughly stabilises at $K=5$. Thus, in the strongly-correlated case
convergence with respect to $K$ is slower, suggesting $K > 4$ for
accurate optimization. Consequently, numerical efforts for energy
optimization exhibit relatively steep scaling, $O(M^5)$ or worse. It
should be taken into account, however, that in this case it would be
more prudent not to try optimization of a large number of generators
at once, but instead combine generators in groups, optimizing the
mostly important (strongly correlated) first and rely on the iterative
\gls{iQCC} procedure. Only in extremely pathological cases---like the
full dissociation of a covalent crystal---a strongly correlated
subsystem will have more than a few dozens of generators.

\begin{figure*}
  \centering
  \includegraphics[width=1.0\textwidth]{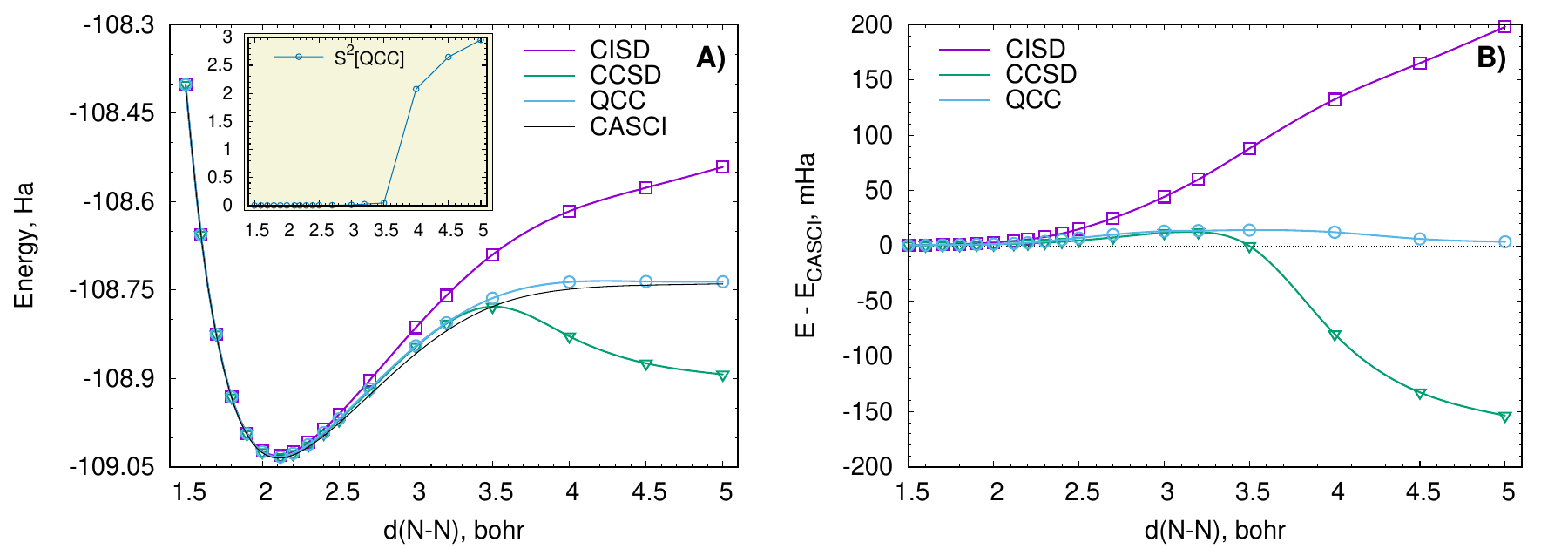}
  \caption{Potential energy curves for \ce{N2}.}
  \label{fig:n2_pec}
\end{figure*}

Finally, we computed the \gls{PEC} for the range \ce{d(N-N)} =
\SIrange{1.5}{5.0}{\bohr} with the full set of potentially non-zero
generators, $M = 136$; see Fig.~\ref{fig:n2_pec}.
Optimization of the \gls{QCC} energy with the exact \gls{QCC}
Ansatz~(\ref{eq:QCC_ans}) with $M = 136$ on a classical computer is
out of the question: the un-truncated $\hat U(\mathbf{t})$ contains
$2^{136} \approx \num{8.7e40}$ terms. Low-order expansions with
$K\le 4$ are tractable, but optimal energies will not be exact.
However, for a 16-qubit system, the dimensionality of the qubit
Hilbert space is no greater than $2^{16} = \num{65536}$. Moreover,
since the second-quantized Hamiltonian also carries information about
states with different number of electrons, spins, \latin{etc.}, its
matrix representation has a block-diagonal form, with blocks that
couple only the states with the same number of electrons (Hamiltonian
is particle-conserving), same spin (we consider systems in the absence
of a magnetic field and ignore spin-orbit coupling), and---if a system
possesses spatial symmetry---belonging to the same irreducible
representation of the corresponding symmetry group. Not all symmetries
can be simply accounted for in the qubit representation. The so-called
qubit tapering technique~\cite{Bravyi:2017/ArXiv/1701.08213,
  Setia:2019/jctc/6091} allows for reducing the dimensionality of
Hilbert space by 5 qubits for systems possessing $D_{\infty h}$
symmetry; hence, the symmetry-reduced Hamiltonian for the state of a
particular symmetry, namely, 10-electron singlet $A_{1g}$ state, will
be characterized by only $16 - 5 = 11$ qubits, and the corresponding
qubit Hilbert space dimensionality is merely $2^{11} = \num{2048}$.
This opens up a possibility to find the exact \gls{QCC} energy using
the functional $F^{[N]}(\mathbf{t})$, Eq.~\eqref{eq:E_N_def}, with
$N = 2048$. Our $F^{[N]}(\mathbf{t})$ implementation (see
Sec.~\ref{sec:impl-fn_t}) automatically detects that the number of
unique $\hat X \ket{0}$ states is no greater than \num{2048}. It is
also notable that the dimensionality of invariant subspaces depends on
the set of generators too. As follows from
Table~\ref{tab:N2_at2.118_nent_22}, for $M = 22$ generators ranked by
Eq.~(\ref{eq:un1_ranking}) the size of the invariant subspace is 1024,
while in the case of stretched molecule and $M = 19$ it is merely 128,
see Table~\ref{tab:N2_at4.0_nent_19}.

Another salient feature of the \gls{QCC} Ansatz is its
order-dependence, which is common among other \gls{VQE}
methods~\cite{Evangelista:2019/jcp/244112, Grimsley:2020/jctc/1,
  Izmaylov:2020/pccp/12980, Hirsbrunner:2024/quantum/0}. This
\emph{ordering problem} has the fundamental consequences, such as
whether the particular ordering guarantees the exact solution or
not~\cite{Evangelista:2019/jcp/244112}, or can significantly impact on
shapes of \glspl{PEC}~\cite{Grimsley:2020/jctc/1}. In the \gls{QCC}
Ansatz the order of generators is determined by the ranking scheme;
see Sec.~\ref{sec:generator-ranking}. Unfortunately, ordering of
generators with numerically close or identical ranking values can
still be affected by uncontrollable numerical noise. We solved this
issue by first assuming that generators are \emph{degenerate} if their
ranking values rounded to \num{e-11} are identical. These degenerate
generators, in turn, were ordered according to the binary
representation of their $\hat X$ factors, see
Eq.~(\ref{eq:pauli_factorization}), thus completely fixing ambiguity.
It must be noted that the ordering problem in the \emph{iterative}
approach (complete \gls{iQCC}) is virtually non-existent: the method
will eventually converge to a \gls{FCI} or \gls{CASCI} value (which is
unique) albeit with slightly different rates.

In Fig.~\ref{fig:n2_pec} we also show \glspl{PEC} computed by
conventional quantum chemistry methods: the \gls{CISD}, \gls{CCSD},
and \gls{CASCI} ones; all of them used the same active space,
CAS(10,~8). For the range of \ce{N-N} distances where correlation is
weak-to-moderate (\SIrange{1.5}{2.5}{\bohr}), both \gls{CCSD} and
\gls{QCC} methods agree very well with the \gls{CASCI} reference and
each other: deviations from the \gls{CASCI} energy do not exceed
\SI{10}{\milli\hartree}. The \gls{CCSD} method is consistently closer
to the \gls{CASCI} reference than \gls{QCC}, but this could be
explained by its non-variational nature. Starting from \ce{d(N-N)} =
\SI{3.5}{\bohr} the quality of estimates given by \gls{CCSD} and,
especially, the \gls{CISD} method quickly deteriorates while the
\gls{QCC} method performs noticeably better: its deviation from
\gls{CASCI} does not exceed \SI{15}{\milli\hartree}. However, the
apparent quality of the \gls{QCC} energies is not confirmed by quality
of underlying wave function: form the inset in Fig.~\ref{fig:n2_pec}A
it is clear that the QCC wave function becomes strongly
spin-contaminated starting from $\ce{d(N-N)} = \SI{3.5}{\bohr}$: the
mean value of $\hat S^2$ operator jumps from essentially zero to
almost 3 at \SI{5.0}{\bohr}. This is another manifestation of
L\"owdin's symmetry dilemma~\cite{Lowdin:1963/rmp/496} but for the
correlated wave function.

In general, the \gls{QCC} method with the full set of generators
provides energies that are considerably better than \gls{CISD} ones
and close to the results of \gls{CCSD} method when correlation is not
too strong. \Gls{QCC} energy estimates also demonstrate robustness to
changes in the electronic structure of a system of interest. Overall,
the \gls{QCC} method is similar to the disentangled (factorized)
\gls{UCC} method~\cite{Evangelista:2019/jcp/244112,
  Chen:2021/jctc/841, Chen:2022/jctc/2193}, which uses different form
of generators.

\subsection{Water molecule (\ce{H_2O}, 36 qubits)}
\label{sec:ceh_2o}

Preparatory calculations were the following. We have performed
\gls{RHF} calculations using Pople's 6-31G(d) basis
set~\cite{Hariharan:1973/tca/213} for a near-equilibrium molecular
configuration, $\ce{d(O-H)} = \SI{0.95}{\angstrom}$. The valence
$\angle \ce{HOH}$ was fixed at \SI{107.6}{\degree}. The \gls{CAS} was
formed by taking all \gls{RHF} \glspl{MO} except the lowest-energy
one, which correlates with core 1s orbital of the \ce{O} atom,
resulting in CAS(8,~18). The 36-qubit Hamiltonian, prepared as
described in Sec.~\ref{sec:gener-char-trial}, contained \num{41907}
terms greater than \num{e-8} in magnitude.

In this example we examined the use of $F^{[N]}(\mathbf{t})$,
Eq.~(\ref{eq:E_N_def}), for energy extrapolation when the exact
\gls{QCC} energy is unattainable. The 36-qubit Hamiltonian has 7929
groups associated with all possible single and double
excitations---this is the total number of generators which satisfy
non-zero gradient criterion, Eq.~(\ref{eq:gradE}). As was already
noted above, some of them (for example, single and double
virtual-virtual vv/vvvv excitations) are associated with numerically
zero gradients for the 8-electron singlet reference state. To simplify
our task, we retained only single and double excitations of ov and
oovv types. Since gradients of the ov-type excitations are very small
because of using the Hartree--Fock orbitals, all included generators
are naturally arranged ``doubles first, singles
last''~\cite{Evangelista:2019/jcp/244112}, leading to the \gls{QCC}
Anzatz with $M = 1182$ generators.

We estimate the dimensionality of the 36-qubit Hilbert space that
contains the exact ground state as $2^{32} = \num{4294967296}$ because
the qubit tapering technique~\cite{Bravyi:2017/ArXiv/1701.08213,
  Setia:2019/jctc/6091} allows for reducing the dimensionality of
Hilbert space by 4 qubits for the system with $C_{2v}$ symmetry. Thus,
the exact solution can be obtained with $F^{[N]}(\mathbf{t})$ for
$N = \num{4294967296}$, which is currently impossible in our
implementation. Instead, we assessed the following scheme: first, we
optimize amplitudes with $E^{[K]}(\mathbf{t})$ for $K = 2$. Despite of
not being fully convergent, they might be a fair approximation to the
exact result, especially for the weakly correlated case. Next, we
calculate a sequence of energies with $F^{[N]}(\mathbf{t})$ using
these amplitudes and values of $N$ ranging from
$2^{18} = \num{262144}$ to $2^{25} = \num{33554432}$ with the ratio of
2; see Table~\ref{tab:FN_conv}.
\begin{table}
  \centering
  \caption{Convergence of $F^{[N]}(\textbf{t})$,
    Eq.~(\ref{eq:E_N_def}), with respect to the size of the expansion
    space $N$. Amplitudes are optimized with $E^{[2]}(\mathbf{t})$
    energy expression, Eq.~(\ref{eq:E_K_def}); the optimized energy
    value is \num{-76.196411}. Timings are collected for 48-thread
    execution on an AMD EPYC 24-core processor with 2~TB
    of RAM.}
  \begin{tabularx}{1.0\linewidth}{XXl}
    \toprule
    N & Energy, \si{\hartree} & Time, \si{\second} \\
    \midrule
    \num{262144 } & -76.\textbf{1962}776  &     \num{213  } \\
    \num{524288 } & -76.\textbf{1962}569  &     \num{559  } \\
    \num{1048576} & -76.\textbf{1962}433  &     \num{1634 } \\
    \num{2097152} & -76.\textbf{1962}345  &     \num{5259 } \\
    \num{4194304} & -76.\textbf{19623}07  &     \num{18315} \\
    \num{8388608} & -76.\textbf{196229}9  &     \num{67353} \\
    \num{16777216}& -76.\textbf{196229}7  &     \num{249700}\\
    \num{33554432}& -76.\textbf{1962296}1&	\num{958630}\\
    \bottomrule
  \end{tabularx}
  \label{tab:FN_conv}
\end{table}
From the results in Table~\ref{tab:FN_conv} it is clear that we have
achieved sub-\si{\micro\hartree} convergence of the total energy
already at $N = 2^{24} = \num{16777216}$, which is only 0.4\% of the
full Hilbert space dimensionality. As was anticipated for the weakly
correlated case, the exact energy estimate is close to the optimized
$E^{[2]}(\mathbf{t})$ energy. Timings demonstrate a transition from a
linear-scaling evaluation of the expansion vector to a quadratic
regime associated with energy evaluation via vector-matrix
multiplication. Expansion vector is computed on a single thread, while
the matrix-vector multiplication can utilize all available cores, so
that using more cores will effectively reduce elapsed (``wall-clock'')
time.

\subsection{Large-scale (80 qubits) iQCC calculations for
  \ce{Ir(F2ppy)3} and energy extrapolation with
  $E^{[1]}(\mathbf{t})$.}
\label{sec:large-scale-iqcc}

\begin{figure}
  \centering \includegraphics[width=0.35\textwidth]{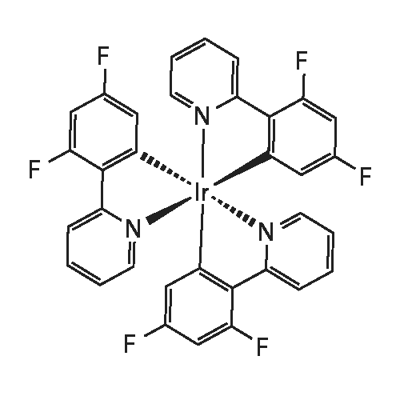}
  \caption{Tris[2-(2,4-difluorophenyl)pyridine]iridium(III)}
  \label{fig:IrF2ppy3}
\end{figure}

To demonstrate scalability of our symmetric polynomial-based optimizer
and integrate it into the \gls{iQCC} workflow, we have performed a
singlet ground-state energy optimization for one of the typical
\gls{OLED} materials,
tris(2-(2,4-difluorophenyl)pyridine)iridium(III), \ce{Ir(F2ppy)3}; see
Fig.~\ref{fig:IrF2ppy3}.
Its 80-qubit Hamiltonian
$\hat H^{(0)}$~\cite{otilumionics/iqcc-hamiltonians}, containing
\num{3532677} terms that are larger than \num{5e-7} in magnitude was
prepared as follows. First, we performed \gls{RHF} calculations on
that molecule using the 6-31G(d) basis set on light atoms (\ce{H},
\ce{C}, \ce{N}, and \ce{F}), and SBKJC pseudopotential + DZ valence
basis set~\cite{Stevens:1992/cjc/612} on \ce{Ir}. 6-component
(Cartesian) $d$ harmonics were used throughout. The converged
Hartree--Fock energy for the lowest singlet $S_0$ state was
\SI{-2124.157531}{\hartree}. 20 occupied \gls{RHF} \glspl{MO} along
with 20 virtual \gls{RHF} orbitals that are closest to the Fermi level
formed CAS(40,~40), from which we generate the qubit expression as
described in Sec.~\ref{sec:simulations}.

\begin{figure}
  \centering \includegraphics[width=0.5\textwidth]{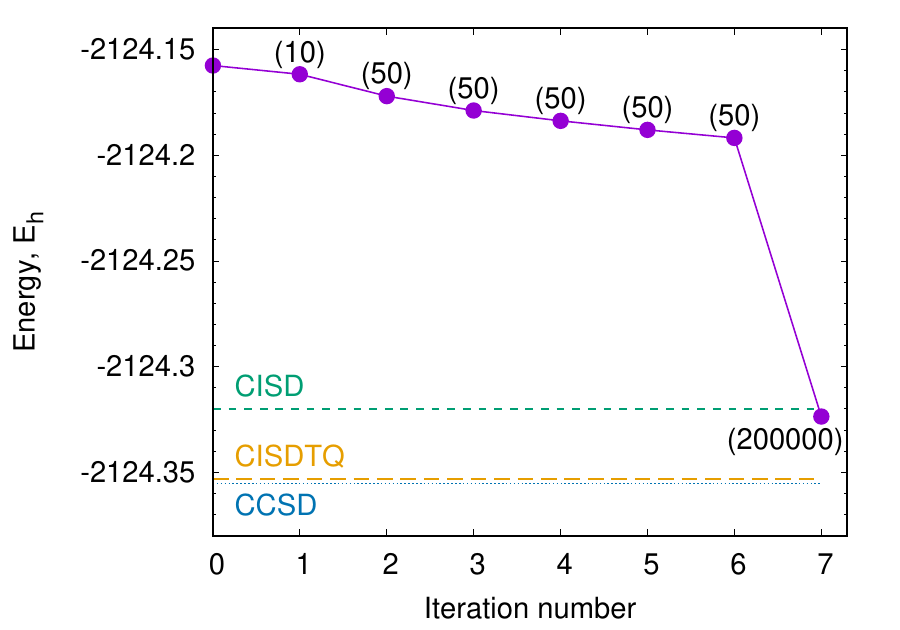}
  \caption{Convergence of \gls{iQCC} iterations for \ce{Ir(F2ppy)3}.
    The zeroth iteration shows the initial \gls{RHF} energy,
    \SI{-2124.157531}{\hartree}. The numbers in parenthesis are the
    number of generators used at the corresponding iteration.}
  \label{fig:irf2ppy3_conv}
\end{figure}

We have executed seven \gls{iQCC} iterations with
\numlist{10;50;50;50;50;50;200000} generators ranked according to the
measure~(\ref{eq:un1_ranking}), optimizing $E^{[K]}(\mathbf{t})$
energy expression with $K = {}$\numlist{10;3;3;3;3;1}, respectively.
All the steps except the last one involved dressing of the current
Hamiltonian by Eq.~(\ref{eq:dressed-H}) and dropping the terms that
are smaller than \num{5e-7} in magnitude. The size of the (dressed)
Hamiltonian at the final 7-th iteration was \num{620228571} terms.
Choosing 10 generators at the first iteration allowed us to include a
group of spin-preserving generators and decrease the maximum ranking
value from \num{0.0611} to \num{0.0453} without spin symmetry
breaking. At all subsequent iterations the wave function becomes
progressively spin-contaminated, with $\Braket{S^2}$ value reaching
\num{0.0287} at the 7-th iteration, while the maximum ranking value
decreased to \num{0.0205}. Our final \gls{iQCC} variational energy
estimate after amplitude optimization with \num{200000} generators was
\SI{-2124.323570}{\hartree}, This value should be compared to the
\gls{CISD}, CISDTQ, and \gls{CCSD} energies computed for the same
active space, see Fig.~\ref{fig:irf2ppy3_conv}.
The \gls{iQCC} method provides energies below the variational
\gls{CISD} estimate, but disconnected quadruple contributions that are
mostly responsible for the difference between \gls{CISD} and
CISDTQ/CCSD estimates (connected triples contributions are small, of
the order of $\approx \SI{0.002}{\hartree}$) are not well-captured at
$K=1$ level. Overall, for this weakly-correlated system the
variational CISDTQ estimate should be rather close to the exact
\gls{CASCI} value.

Timings for the amplitude optimization are shown in
Table~\ref{tab:IrF2ppy3_ampl_opt_timings}.
\begin{table}[!b]
  \centering
  \caption{Timings for the singlet ground-state energy minimization
    for \ce{Ir(F2ppy)3}. 24-thread execution on an AMD EPYC 24-core
    processor with 2~Tb of RAM.}
  \begin{tabularx}{1.0\linewidth}{lX}
    \toprule
    Iteration \textbackslash & Hessian + amplitude optimization, \si{\second} \\
    \midrule
    1&   $3.1 + 2.5    = \num{5.6  }$\footnotemark[1] \\ 
    2&   $2.3 + 0.5    = \num{2.8  }$ \\
    3&   $10.1 + 1.6   = \num{11.7 }$ \\
    4&   $16.8 + 0.1   = \num{16.9 }$ \\
    5&   $25.2 + 6.5   = \num{31.7 }$ \\
    6&   $38.1 + 10.1  = \num{48.2 }$ \\
    7&   $3509 + 80970 = \num{84479}$ \\
    \bottomrule
  \end{tabularx}
  \footnotetext[1]{Including time for \textsc{julia} JIT
    compilation~\cite{Julia:2017}.}
  \label{tab:IrF2ppy3_ampl_opt_timings}
\end{table}
It is clear that even with a large number of generators, timings for
the amplitude optimization with our new optimizer are rather modest.

We also compare our optimizer with Amazon's sparse wave function
simulator (AWS)~\cite{Steiger:2024/arXiv/2404.10047}. AWS used
\texttt{r7i.metal-48xl} Amazon EC2 instance. An instance is a
two-socket computer with Intel(R) Xeon(R) Platinum 8488C CPUs, each
with 48 cores and two threads per core. Timings as well as other
relevant quantities are summarized in
Table~\ref{tab:Amazon_comparison}.
\begin{table}
  \centering
  \caption{Comparison of the current energy/amplitude optimizer with
    with Amazon's sparse wave function simulator
    (AWS)~\cite{Steiger:2024/arXiv/2404.10047}.}
  \begin{tabularx}{1.0\linewidth}{@{}Xll}
    \toprule
    & $E^{[K]}(\mathbf{t})$ & AWS \\
    \midrule
    Final energy, \si{\hartree}         & \num{-2124.323570} & \num{-2124.225246} \\
    Generators used                     & \num{200110}       &       \num{1100}   \\
    Total simulation time, \si{\second} & \num{119221}       &       \num{56071}  \\
    Number of CPUs                      & \num{24}           &       \num{96}    \\
    CPU time, \si{\hour}                & \num{563.2}        &       \num{1495.2} \\ 
    \bottomrule
  \end{tabularx}
  \label{tab:Amazon_comparison}
\end{table}

We would like to emphasize that the total time reported for the
\gls{iQCC} method contains many other steps that are not performed by
the AWS; the corresponding breakdown of the total time is given in
Table~\ref{tab:timings_breakdown}.
\begin{table}[!b]
  \centering
  \caption{The total simulation time breakdown for the \gls{iQCC}
    method; only rate-limiting steps are shown.}
  \begin{tabularx}{1.0\linewidth}{@{}Xl}
    \toprule
    Step & Time, \si{\second} \\
    \midrule
    Hamiltonian dressing  & \num{27554}  \\
    Hessian + amplitude optimization & \num{84596} \\
    \bfseries Total\footnotemark[1] & \num{112151} \\ 
    \bottomrule
  \end{tabularx}
  \footnotetext[1]{Auxiliary calculations contribute to the difference
    with the total execution time reported in
    Table~\ref{tab:IrF2ppy3_ampl_opt_timings}.}
  \label{tab:timings_breakdown}
\end{table}

\section{Conclusions}
\label{sec:conclusions}

The \gls{QCC} Ansatz~(\ref{eq:QCC_ans}) with $M$ generators that lies
at the heart of the \gls{iQCC} method is exponentially hard for
classical computers. To make it more amicable to classical hardware,
we propose two approximations for the \gls{QCC} energy
expression~(\ref{eq:qcc_energy}): $E^{[K]}(\mathbf{t})$,
Eq.~(\ref{eq:E_K_def}) and $F^{[N]}(\mathbf{t})$,
Eq.~(\ref{eq:E_N_def}). The first one, $E^{[K]}(\mathbf{t})$, is based
on the symmetric polynomial expansion of the exact expression up to
the specified order $1 \le K \le M$. We have demonstrated that the
optimal energy and amplitudes obtained with $E^{[K]}(\mathbf{t})$
converge geometrically with increasing $K$. The onset of the
geometrical convergence depends on a system under consideration: for
weakly correlated systems chemically accurate energies can be obtained
with $K=2$ and microhartree accuracy can be achieved already at
$K = 3$. For strongly correlated systems the convergence is somewhat
delayed, but microhartree accuracy can be obtained at $K = 6$. The
value $K = 4$ is recommended for expedite calculations with the
chemical accuracy ($\le \SI{1}{\milli\hartree}$) in the case of strong
correlation. For any $K$, $E^{[K]}(\mathbf{t})$ is a differentiable
expression with continuous gradients which are necessary for applying
efficient gradient-based optimization algorithms, such as
\textsc{l-bfgs}~\cite{lbfgs-scipy-Byrd:1995/jsc/1190}. Remarkably,
$E^{[K]}(\mathbf{t})$ for $K = 1$ is equivalent to a linear
parametrisation of a correlated wave function, known as the \gls{CI}
expansion in the traditional quantum chemistry with the mapping of
\gls{CI} expansion coefficients to cluster amplitudes via
Eq.~(\ref{eq:C_def}). \Gls{CI}-like expansion can be optimized with
hundreds of thousands of generators, which we exploited to extrapolate
the energies to the \gls{CASCI} limit, see
Sec.~\ref{sec:large-scale-iqcc}.

Continue along the line of simplification, we have introduced the
\acrfull{DHA} (see Sec.~\ref{sec:diag-hess-appr}) that is based on
$E^{[1]}(\mathbf{t})$, which can be used not only as an extrapolation
technique allowing for including hundreds of millions generators, but
also for ranking them, thus defining an alternative to ``first-order
unitary ranking'' discussed in Sec.~\ref{sec:generator-ranking}.

The main obstacle with $E^{[K]}(\mathbf{t})$ is steep scaling of
computational efforts with increasing $M$ and $K$ as $O(M^K)$ for
$K \ll M$; see Eq.~(\ref{eq:N_K_for_E_K}) and
Table~\ref{tab:U_K_size}. To address this issue we introduced our
second energy expression $F^{[N]}(\mathbf{t})$,
Eq.~(\ref{eq:E_N_def}), where $N$ is the dimensionality of the
expansion space---in other words, the number of ``qubit exited
states''. The latter---the number of qubit product states for $n$
qubits---is fundamentally limited by the size of the corresponding
Hilbert space, $2^{n}$, but in the most of practically relevant cases
$n \gg 30$ ($ \sim 15$ active orbitals), and the limit is
unattainable. However, as is established in the \gls{CI}
theory~\cite{Ivanic:2001/tca/339}, much fewer states may be needed for
chemically accurate total energies. To test this assertion, we
performed a convergence study of $F^{[N]}(\mathbf{t})$ for increasing
$N$. In the particular case of 36-qubit \ce{H2O} Hamiltonian, for
which we estimate the size of the relevant Hilbert space as
$N_\text{lim} = 2^{32}$, $F^{[N]}(\mathbf{t})$ energies for
$N = 2^{24} \approx \SI{0.4}{\percent}$ of $N_\text{lim}$ appear to
converge to $\si{\micro\hartree}$ accuracy. Despite its flexibility
and computational efficiency, $F^{[N]}(\mathbf{t})$ is less suitable
for optimization, because it encodes, in general, a non-smooth
function of amplitudes as the composition of the expansion space can
drastically change even for close but distinct amplitude vectors
$\mathbf{t}$.

In the first time we have calculated the entire \gls{PEC} for the
\ce{N2} molecule using the \gls{QCC} Ansatz with all possible
generators satisfying the non-zero gradient
criterion~(\ref{eq:gradE}). Our results suggest that the \gls{QCC}
form is an alternative to the disentangled/factorized form of the
\gls{UCC} method~\cite{Evangelista:2019/jcp/244112,
  Chen:2021/jctc/841, Halder:2022/jcp/174117, Chen:2022/jctc/2193}
whose generators are derived from fermion single- and
double-excitation operators. An important advantage of \gls{QCC}
generators is that they not need to be derived from any fermionic
predecessors, which ultimately leads to the iterative scheme realized
as the \acrfull{iQCC} method~\cite{Lang:2020/arXiv/2002.05701}.

Fundamentally, we have demonstrated that the maximum possible
entanglement that may be provided by a noise-free quantum computer is
hardly useful in molecular applications. On the contrary, relatively
low degree of correlation, expressed by low values of $K$, are
sufficient to provide reliable and useful-for-chemistry ground-state
energy estimates. Although this conclusion may not be correct for
\emph{excited-state} calculations, it is likely valid for the
singlet-triplet gap calculations that are relevant to designing new
phosphorescent materials for the \gls{OLED} industry.

In perspective, it is interesting to test our developments of a wider
class of molecules, especially containing large open-shell subsystems
(``molecular magnets''), larger industrially relevant molecules, like
phosphorescent and fluorescent transition-metal complexes,
transition-metal catalysts~\cite{vonBurg:2021/prr/033055}, and other
challenging systems.

\appendix

\section{Implementation of $E^{[K]}(\mathbf{t})$}
\label{sec:impl-ek_t}

Equation~\eqref{eq:E_K_def} is evaluated in two stages. First, all
intermediates that are independent from amplitudes are computed and
stored. We call this stage a \textit{compilation} stage. Second, at a
\emph{computation} stage energies and gradients from provided
amplitudes are evaluated; they can be subsequently used in
gradient-based optimisation methods like
\textsc{l-bfgs}~\cite{lbfgs-scipy-Zhu:1997/atms/550}.

The \textbf{compilation} stage consists of the following steps:
\begin{enumerate}
\item
  \label{item:c1}
  From an ordered list of operators $(-\I \hat T_k)$~\footnote{It is
    convenient to include a factor $(-\I)$ into the definition of
    generators.} that define a \gls{QCC} Ansatz, generate a list of
  terms of symmetric polynomials of the orders from 0 to $K$. The
  total number of terms in the expansion is given by
  Eq.~\eqref{eq:N_K_for_E_K}.
  
\item
  \label{item:c2}
  Every term in the list above is a Pauli word $\hat T_{\bar q}$ with
  a unimodular coefficient $D_{\bar q} = \pm 1$ or $\pm \I$. We
  factorize $\hat T_{\bar q}$ as in Eq.~\eqref{eq:T_ket0} into a list
  of phases $\phi_{\bar q}$ and lists of $\hat X_{\bar q}$ and
  $\hat Z_{\bar q}$ operators. A list of $Z_{\bar q}$ operators is
  immediately consumed to evaluate $f_{\bar q}$ [see
  Eq.~\eqref{eq:T_ket0}], which are merged with the phases
  $\phi_{\bar q}$ and the corresponding coefficients $D_{\bar q}$. The
  resulting phases must be real $\pm 1$; imaginary values signal the
  error in definition of generators~\footnote{Generators $\hat T_k$ of
    the \gls{QCC} unitary must be chosen so as to contain the
    \emph{odd} number of $\hat y$ Pauli elementary operators to ensure
    they are \emph{imaginary} (\latin{i.e.} have purely imaginary
    matrix elements in the computational basis), or, equivalently
    $-\I \hat T_k$ are \emph{real} operators.} or in the coding.

\item
  \label{item:c3}
  A list of $\hat X_{\bar q}$ operators is sorted and unique terms are
  identified and indexed. A list of lists with positions of the unique
  $\hat X_q$ operators is created.

\item
  \label{item:c4}
  For the list of unique $\hat X_q$ operators compute and store the
  Hessian matrix $\mathbf{H}$ with elements
  $\mathbf{H}_{q'q} = \braket{0|\hat X_{q'} \hat H \hat X_{q} |0}$.
  This matrix can also be sparse, so a sparse storage is used to
  further reduce memory consumption.

\item
  \label{item:c5}
  Normalization factor
  $\mathbf{N} = \braket{0|\hat U^\dagger(\mathbf{t}) \hat
    U(\mathbf{t})|0}$ has unique matrix elements
  $\mathbf{N}_{q'q} = \braket{0|\hat X_{q'} \hat X_q|0} =
  \delta_{q'q}$

\end{enumerate}

The worst-case complexity of the compilation stage is determined by
evaluation of the matrix $\mathbf{H}$ at the step~\ref{item:c4}. If
the maximum possible order $K = M$ (the exact energy functional) is
selected and all vectors in the expansion of
$\hat U(\mathbf{t})\ket{0}$ are linearly independent, one needs to
evaluate $O(4^M)$ matrix elements. For the $K \ll M$ the length of the
expansion is $O(M^K)$ and one needs $O(M^{2K})$ matrix elements. The
algorithm complexity is, however, linear in the number of terms in the
qubit Hamiltonian with a very small prefactor, which allows for either
large systems or many \gls{iQCC} iterations. Terms coalescence
(redundancy) significantly affects the complexity as the size
$\mathbf{H}$ can be much smaller than the length of the
expansion~\eqref{eq:qcc_expanded}, but savings are difficult to
quantify \latin{a priori} as they are system- and generator
list-dependent.

At the \textbf{computation} stage trigonometric functions of
amplitudes are evaluated first. The remaining steps are:
\begin{enumerate}
\item
  \label{item:e1}
  From the list of $\tan{\left(t_k/2\right)}$ the list of terms of the
  symmetric polynomials in $\tan{\left(t_k/2\right)}$ from order $k=0$
  to $K$ are generated by the same routine as at the
  step~\ref{item:c1} of the compilation stage. This list is
  element-wise updated by overall phases computed at the compilation
  stage and scaled by a product of cosines of amplitudes,
  $\prod_{k=1}^K \cos{\left(t_k/2\right)}$. As a result, a vector
  $\mathbf{U}$ representing $\hat U(\mathbf{t})\ket{0}$ for a given
  set of amplitudes $\mathbf{t}$ is obtained.

\item
  \label{item:e2}
  The vector $\mathbf{U}$ from the previous step is multiplied by the
  Hessian matrix $\mathbf{H}$ from the left to produce a resulting
  vector $\mathbf{HU}$ which is stored separately. In the absence of
  redundancy this is tantamount to a simple matrix-vector product.
  Redundancy adds an additional loop structure: one needs to go over
  all \textit{unique} $\hat X$ terms and for each of them calculate a
  sum across all occurrence of that term in $\mathbf{U}$. To calculate
  energy, a scalar product of $\mathbf{U}$ with $\mathbf{HU}$ is taken
  and if only energies are requested, explicit storage of
  $\mathbf{HU}$ can be avoided. This vector, however, is used in
  gradient calculations, see Appendix~\ref{sec:computing-gradients}.
\end{enumerate}
Finally, computed energies and gradients are returned to a caller.
Computational complexity of that stage is determined by calculations
of the vector $\mathbf{HU}$. The computation stage is invoked each
time when amplitudes are updated, for example, by iterative
optimization.

\section{Implementation of $F^{[N]}(\mathbf{t})$}
\label{sec:impl-fn_t}

Evaluation of $F^{[N]}(\mathbf{t})$ is done as follows. We iterate by
integer $k$ from 0 to $(M-1)$. Let $\mathbf{c}_k$ be a real array of
coefficients that depend on amplitudes and phases (introduced below),
and an array $\mathbf{w}_k$ which holds the binary representation of
the basis product states. The basis-state array $\mathbf{w}_k$ is
stored \emph{sorted} according to the natural binary order of
elements; ordering of coefficients is always adjusted accordingly.

Iterations are initialized by setting $\mathbf{c}_0 =1$ and
$\mathbf{w}_0 = \text{binary rep}\ket{0}$.

At the expansion step we perform the following operations:
\begin{enumerate}
\item Create a new vector of coefficients
  $\tilde{\mathbf{c}}_{k+1} = \sin(t_{M-k}) \mathbf{c}_k$.

\item Update the current $\mathbf{c}_k$ by $\cos(t_{M-k})$.

\item Factorize the current operator $-\I \hat T_{M-k}$ into a phase
  $\phi_{M-k}$ and operators $\hat X_{M-k}$ and $\hat Z_{M-k}$ that
  satisfy
  \begin{equation}
    \label{eq:op_ising_left_decomp}
    -\I \hat T_{M-k} = \phi_{M-k} \hat X_{M-k} \hat Z_{M-k}
  \end{equation}
  Note that for the properly chosen generators $\hat T_k$ the phases
  $\phi_k$ come out real ($\pm 1$). The phase $\phi_{M-k}$ is used to
  update $\tilde{\mathbf{c}}_{k+1}$:
  \begin{equation}
    \label{eq:1st_phase_update}
    \tilde{\mathbf{c}}_{k+1} = \phi_{M-k}\tilde{\mathbf{c}}_{k+1}
  \end{equation}
  
\item Evaluate a phase vector $\mathbf{f}_{M-k}$ by action of the
  operator $\hat Z_{M-k}$ onto the current basis states vector
  $\mathbf{w}_k$:
  \begin{equation}
    \label{eq:Z_action}
    \hat Z_{M-k} \mathbf{w}_k = \mathbf{f}_{M-k} \mathbf{w}_k. 
  \end{equation}
  Note that one can avoid explicit storage of $\mathbf{f}_{M-k}$
  because the resulting phases $\pm 1$ can be immediately combined
  (merged) with elements of $\tilde{\mathbf{c}}_{k+1}$.

\item Create a new basis state vector $\tilde{\mathbf{w}}_{k+1}$ by
  action of $\hat X_{M-k}$ on $\mathbf{w}_k$:
  \begin{equation}
    \label{eq:X_action}
    \tilde{\mathbf{w}}_{k+1} = \hat X_{M-k} \mathbf{w}_k.
  \end{equation}
  Note that elements of $\tilde{\mathbf{w}}_{k+1}$ representing the
  qubit product states are no longer sorted but unique. We sort
  $\tilde{\mathbf{w}}_{k+1}$ according to the same ordering as for
  $\mathbf{w}_k$. We also reorder elements of
  $\tilde{\mathbf{c}}_{k+1}$ to match the (new) order of elements in
  the sorted $\tilde{\mathbf{w}}_{k+1}$.
  
\item Now we can \emph{merge} two sorted basis set vectors,
  $\mathbf{w}_k$ and $\tilde{\mathbf{w}}_{k+1}$ into the final
  $\mathbf{w}_{k+1}$. During this process, which is similar to the
  merge stage of the merge sort algorithm, we can identify the
  identical elements in both $\mathbf{w}_k$ and $\tilde{\mathbf{w}}_k$
  and sum up the corresponding coefficients. As a result, the final
  $\mathbf{c}_{k+1}$ is produced.
\end{enumerate}
The expansion stage is done. If the length of $\mathbf{w}_{k+1}$ is
less or equal to $N$ we can continue with the next iteration.
Otherwise, we proceed with the \emph{compression (truncation)} stage.

At the compression stage one needs to find a permutation $p$ the sorts
$\mathbf{c}_{k+1}$ in descending order of absolute values. Then we
compute a norm of a ``residual'' vector
$\mathbf{c}_{k+1}[p[N+1:\text{end}]]$ (we use the
\textsc{julia}~\cite{Julia:2017} notation for indexing vectors) to
evaluate a re-normalization coefficient $\lambda_k$.

Afterwards, we truncate the arrays $\mathbf{w}_{k+1}$ and
$\mathbf{c}_{k+1}$ to $N$ via permutation $p$ as
\begin{align}
  \label{eq:w_trunc_N}
  \mathbf{w}_{k+1} & = \mathbf{w}_{k+1}[p[1:N]], \\
  \label{eq:c_trunc_N}
  \mathbf{c}_{k+1} & = \mathbf{c}_{k+1}[p[1:N]].
\end{align}
Additionally, the new $\mathbf{c}_{k+1}$ is multiplied by
$\lambda_k^{-1}$ to restore normalization. The $k$-th iteration is
complete and one can proceed until $k=M$.

Finally, one has to compute the matrix elements
$\braket{w_{j'}|\hat H| w_j}$, $1 \le j',j \le N$ to generate a
Hessian matrix $\mathbf{H}$ (see Sec.~\ref{sec:symm-polyn-expans}
where $\hat X_{q}\ket{0}$ play a role of $\ket{w_j}$). The energy
$F^{[N]}(\mathbf{t})$ is computed as a matrix-vector product
\begin{equation}
  \label{eq:FN_energy}
  F(\mathbf{t}) = \mathbf{c}_M^\dagger \mathbf{H} \mathbf{c}_M. 
\end{equation}
Note that it is not necessary to store the entire matrix $\mathbf{H}$
explicitly. Instead, one can generate a single row $\mathbf{H}_{j*}$
and multiply it by $\mathbf{c}_M$ to generate an element
$(\mathbf{Hc}_{M})_j$. These calculations can be done in parallel on
multiple workers assigned to different CPU cores on a shared-memory
machine or different processes on distributed-memory architectures, so
that the process is expected to be scalable to any numbers of
available workers.

\section{Computing gradients}
\label{sec:computing-gradients}

We commence from the definition of \gls{QCC} energy,
Eq.~\eqref{eq:qcc_energy}. The partial derivative of it with respect
to $t_k$ amplitude is
\begin{equation}
  \label{eq:qcc_de_dtk}
  \frac{\partial E}{\partial t_k} = \Braket{0|
    \frac{\partial {\hat U^\dagger(\mathbf{t})}}{\partial t_k}
    \hat H  \hat U(\mathbf{t}) |0} + \text{c.c}   
\end{equation}
where c.c stands for complex conjugation. If $\hat U(\mathbf{t})$ is
real, which is the case, two terms are equal and we can use the first
one as our next step.

Equation~\eqref{eq:qcc_de_dtk} can be interpreted as a scalar product
of vectors
$\mathbf{dU}_k = \frac{\partial {\hat U(\mathbf{t})}}{\partial t_k}
\ket{0}$ with the vector $\mathbf{HU}$ (see
Sec.~\ref{sec:symm-polyn-expans}) which is readily available during
energy evaluation.

From the definition of the \gls{QCC} Ansatz as
\begin{equation}
  \label{eq:qcc_ansatz_alt}
  \hat U(\mathbf{t}) = \prod_{k=1}^M \left(\cos(t_k/2) -
    \I\sin(t_k/2) \hat T_k \right) 
\end{equation}
it is clear that taking partial derivative of the
expression~\eqref{eq:qcc_ansatz_alt} with respect to $t_k$ amounts to
replacing the corresponding cos/sin functions in parentheses with
their derivatives, namely,
$1/2\left(-\sin(t_k/2) - \I\cos(t_k/2)\hat T_k \right)$. That is, at
the computation stage one needs to appropriately modify a list of
trigonometric functions and the rest of algorithm runs unimpeded to
yield a vector $\mathbf{dU}_k$. There is one new vector per every
gradient component, so $M$ additional vectors are generated. These
vectors are scalar-multiplied to $\mathbf{HU}$ giving an $M$-component
gradient vector. A computational overhead of evaluating of a gradient
vector is small compared to evaluation of $\mathbf{HU}$ itself, which
is required by energy calculations anyway. Thus, the suggested
algorithm is computationally more efficient than other methods of
evaluating gradients, such as the parameter-shift
rule~\cite{Schuld:2019/pra/032331, Kottmann:2021/cs/3497_si,
  Izmaylov:2021/pra/062443}---in this case the gradient complexity is
strictly linear in $M$.

\bibliographystyle{apsrev4-2}

\begin{thebibliography}{82}
\bibitem [{Noteb()}]{Noteb}%
  \BibitemOpen
  \bibinfo {note} {We use lowercase bold letters to denote vectors in $2^n$
  qubit Hilbert space and capital bold letters to define subspaces where those
  vectors live in.}\BibitemShut {Stop}%
\bibitem [{Notec()}]{Notec}%
  \BibitemOpen
  \bibinfo {note} {It is convenient to include a factor $(-\protect \mathrm
  {i}\mkern 1mu)$ into the definition of generators.}\BibitemShut {Stop}%
\end{thebibliography}

\end{document}